\documentclass[12pt]{article}
\usepackage{comment}
\usepackage[margin=2cm]{geometry}
\usepackage{amsmath,amssymb,extarrows,mathtools,graphicx,subfigure,setspace,bbold}
\usepackage{cite}
\usepackage{slashed}
\usepackage[dvipsnames]{xcolor}
\usepackage{hyperref}
\hypersetup{colorlinks=true, linkcolor=blue, citecolor=magenta, linktoc=page}
\usepackage{tikz}
\usepackage{arydshln,leftidx,mathtools}
\usetikzlibrary{decorations.pathreplacing}
\numberwithin{equation}{section}
\newcommand{\be}{\begin{equation}}
\newcommand{\bea}{\begin{eqnarray}}
\newcommand{\eea}{\end{eqnarray}}
\newcommand{\ba}{\begin{align}}
\newcommand{\ea}{\end{align}}
\newcommand{\ee}{\end{equation}}

\usepackage{chngcntr}
\counterwithout*{figure}{section}
\begin{document}

\begin{titlepage}
\thispagestyle{empty}

\begin{flushright}
MPP-2018-282\\
IPM/P-2018/079\\
\end{flushright}

\vspace{.4cm}
\begin{center}
\noindent{\Large \textbf{Entanglement Evolution in Lifshitz-type Scalar Theories}}\\

\vspace{2cm}

M. Reza Mohammadi Mozaffar$^{a,b}$
and Ali Mollabashi$^{c}$
\vspace{1cm}

{\it $a$ Department of Physics, University of Guilan,
\\
P.O. Box 41335-1914, Rasht, Iran}
\\
\vspace{.6cm}
{\it $b$ School of Physics, Institute for Research in Fundamental Sciences (IPM),}
\\
{\it 19538-33511, Tehran, Iran}
\\
\vspace{.6cm}
{\it $c$ Max-Planck-Institut f\"ur Physik, Werner-Heisenberg-Institut}
\\
{\it 80805 M\"unchen, Germany}
\\
\vspace{1cm}
Emails: {\tt mmohammadi@guilan.ac.ir, alim@mpp.mpg.de}

\vskip 2em
\end{center}

\vspace{.5cm}
\begin{abstract}
We study propagation of entanglement after a mass quench in free scalar Lifshitz theories. We show that entanglement entropy goes across three distinct growth regimes before relaxing to a generalized Gibbs ensemble, namely, \textit{initial rapid growth}, \textit{main linear growth} and \textit{tortoise saturation}. We show that although a wide spectrum of quasi-particles are responsible for entanglement propagation, as long as the occupation number of the zero mode is not divergent, the linear main growth regime is dominated by the fastest quasi-particle propagating on the edges of a \textit{widen} light-cone. We present strong evidences in support of effective causality and therefore define an effective notion of \textit{saturation time} in these theories. The larger the dynamical exponent is, the shorter the linear main growth regime becomes. Due to a pile of tortoise modes which become dominant after saturation of fast modes, exact saturation time is postponed to infinity.

\end{abstract}
\end{titlepage}

\newpage
\tableofcontents
\noindent
\hrulefill
\onehalfspacing
\section{Introduction}
Propagation of entanglement in systems with a large number of degrees of freedom is of great importance to understand non-equilibrium physics in such systems \cite{Calabrese:2005in, Alba:2017lpnas}. This has been widely studied via considering a global quench and focusing on the evolution of a given initial state. The evolution of such a system due to the quench is generally a thermalization process. A typical measure for this thermalization process is how the reduced density matrix corresponding to a typical spatial subregion is close to a thermal density matrix.

The evolution of the system after the quench on the other hand is also an equilibration process. Since the post-quench evolution is a unitary evolution, the system will only reach local equilibrium though never a global one. In this also the relevant quantity is a density matrix corresponding to a subsystem which is expected to reach local equilibrium.
At the end of the thermalization process when the system has reached local equilibrium, the local observables are given by the thermal (Gibbs) ensemble.

The story is different in case of integrable models (including free systems). In these systems the observables are much more constrained by an infinite number of conserved charges and the systems does not thermalize ending up in a Gibbs ensemble but relaxes to generalized Gibbs ensemble \cite{GGE, Sotiriadis:2014uza}.    

A typical measure to quantify the evolution of the system in a pure state after a global quench is to study entanglement entropy of a subsystem in the post-quench state. The entanglement entropy is defined as
$$S_A=-\text{Tr}\left[\rho_A\log\rho_A\right]$$
and $\rho_A:=\text{Tr}_{\bar{A}}|\psi\rangle\langle\psi|$ where $|\psi\rangle=e^{-iHt}|\psi_0\rangle$. $|\psi_0\rangle$ is the pure pre-quench state and $H$ is the post-quench Hamiltonian. Here by quench we mean a global quench which is defined by a sudden change of a parameter in the Hamiltonian of the system at $t=0$.

The dynamics of the system and specifically the spread of entanglement in such systems has been widely studied in a large number of papers (see \cite{Calabrese:2005in} and \cite{Alba:2017lvc} and references therein). 
The quasi-particle picture is a core of understanding how entanglement spreads over the system after a global quench. More recently this picture has been improved in \cite{Alba:2017lvc} which makes it valid in a wider scope. 

From a field theoretic point of view, the quasi-particle picture successfully describes propagation of entanglement in  relativistic field theories. This is strongly based on the notion of causality in these theories. Of course the question of entanglement propagation as a probe to study how the system relaxes to a generalized Gibbs ensemble is a very important question in non-relativistic theories. The propagation of entanglement is specifically intertwined with the causality structure together with the equilibration process of the theory under consideration.  
  
The goal of this paper is to investigate how entanglement propagates in a field theory with Lifshitz scaling symmetry  \cite{Lifshitz, Hertz:1976zz} given by
\bea\label{lifscaling}
t\to\lambda^z t\;\;\;\;\;,\;\;\;\;\;\vec{x}\to\lambda \vec{x},
\eea
where $z$ is the dynamical exponent and determines the anisotropy between space and time. This kind of scaling symmetry is typical at critical points where a continuous phase transition takes place.\footnote{Various properties of entanglement entropy is such theories in static cases has been previously studied in \cite{Solodukhin:2009sk, Zhou:2016ykv, Kusuki:2017jxh, MohammadiMozaffar:2017nri, He:2017wla, MohammadiMozaffar:2017chk, Gentle:2017ywk}.} We study evolution of entanglement in the vacuum state of a translational invariant system and try to understand the corresponding physics basically by means of a zero mode analysis.  

The rest of the paper is organized in the following order: As the rest of the introduction section we introduce a discrete version of Lifshitz scalar field theory which we will utilize in our numerical analysis. In section \ref{sec:EEz1} we review propagation of entanglement in relativistic scalar theory after a global quench. In section \ref{sec:EEz} we give analytical arguments for quasi-particle picture in these theories followed by a numerical study of evolution of entanglement entropy. We analyse our numerical studies focusing on the spectrum of these theories, the key role of tortoise modes and the quasi-particle picture.   

\subsection{Lifshitz-type QFTs on Square Lattice}
A quantum field theory that is invariant under a Lifshitz scaling transformation, introduced in Eq.\eqref{lifscaling}, is what we call a Lifshitz-type QFT. As the simplest example we consider a free scalar field in $d+1$-dimensions with the following action\cite{Alexandre:2011kr}
\bea\label{action}
S=\frac{1}{2}\int dt d\vec{x} \left[\dot{\phi}^2-\sum_{i=1}^{d}(\partial_i^z \phi)^2-m^{2z} \phi^2\right].
\eea
The above action has a Lifshitz scaling symmetry in the massless limit $(m=0)$. Although in this manuscript we will study both $m=0$ and $m>0$ theories. This is clearly a generalization of Klein-Gordon theory ($z=1$) to generic $z$ with non-relativistic scaling symmetry. 
The realization of Klein-Gordon theory on a lattice is the well-known ``harmonic lattice". There also exists a family of models which is generalization of harmonic model on a (hyper)square lattice in generic spatial dimensions. These models are known as \textit{Lifshitz harmonic models}\cite{MohammadiMozaffar:2017nri, He:2017wla} and realize the action given in Eq.\eqref{action} on a square lattice.

To have a intuitive picture of the model lets focus on $1+1$-dimensions. In this case the model is a chain of coupled harmonic oscillators, which each oscillator is coupled to $z$ other oscillators from each side where $z$ is the dynamical exponent. Obviously at this level $z$ should be considered as a positive integer.

The Hamiltonian of these models is given as follows
\be\label{eq:LifH}
H=\sum_{n=1}^{N}\left[\frac{p_n^2}{2M}+\frac{M m^{2z}}{2}q_n^2+\frac{K}{2}\left(\sum_{k=0}^z(-1)^{z+k}{{z}\choose{k}} q_{n-1+k}\right)^2 \right].
\ee 
One can easily check that this Hamiltonian reduces to the well-known harmonic model for the case of $z=1$. It is straightforward to generalize this Hamiltonian to higher dimensions which we are not interested in this paper (see \cite{MohammadiMozaffar:2017nri} for 3$d$ generalization). One can also show that after a canonical transformation
$$(q_n,p_n)\to(\sqrt{MK}q_n,p_n/\sqrt{MK}),$$
this is a discretized version of a free Lifshitz theory on a square lattice with mass $m$ and lattice spacing $\epsilon=\sqrt{M/K}$,\footnote{In what follows for simplicity we choose $K=M=1$ without loss of generality.} introduced in Eq.\eqref{action}.

The Hamiltonian of this model in any number of dimensions on a (hyper)square lattice (more precisely on a $d$ dimensional torus) can be diagonalized in a standard way which we are not going to review here leading to the following dispersion relation \cite{MohammadiMozaffar:2017nri, He:2017wla}
\bea\label{dispersion}
\omega^2_{\mathbf{k}}=m^{2z}+\sum_{i=1}^d\left(2\sin\frac{\pi k_{i}}{N_{x_i}}\right)^{2z},
\eea
where $\mathbf{k}=\{k_1,k_2,\cdots,k_d\}$ is the momentum vector, $k_{i}$'s refers to the momentum components in all spatial directions and $N_{x_i}$ refers to the number of sites in the $i$-th spatial direction.

Different aspects of quantum entanglement in the vacuum state of these models has been studied previously \cite{MohammadiMozaffar:2017nri, He:2017wla,MohammadiMozaffar:2017chk}\footnote{Also  \cite{Gentle:2017ywk} studied the entanglement entropy in Lifshitz theory using a holographic setup.}.
We will utilize the covariant correlator method to study entanglement propagation in these models. The method is briefly reviewed in the appendix section.	

\section{Entanglement Propagation in Relativistic Theories}\label{sec:EEz1}
Before getting into the question of how the relaxation of these systems is affected by non-Lorentzian dynamical exponent, in this section we briefly review the corresponding physics in relativistic field theories. The issue has been studied for 2$d$ CFTs in a series of papers starting by \cite{Calabrese:2005in}. Lets first focus on the simplest case which is a global quantum quenches in the context of 2$d$ CFTs. The post-quench Hamiltonian is scale invariant while the pre-quench one is not. Thanks to the simplification due to the conformal symmetry of the post-quench Hamiltonian, EE can be worked out analytically as a function of time. It is well-known that EE exhibits a linear growth behavior up to a saturation time when a sudden transition to a saturation regime happens. The saturation time is proportional to the length of the entangling region (with a proportionality factor of $1/2$) and the saturation value of EE depends on the details of the initial state. These universal features have been verified in various scale invariant models including transverse Ising spin chain in the same reference above. 

\begin{figure}
\begin{center}
\includegraphics[scale=0.4]{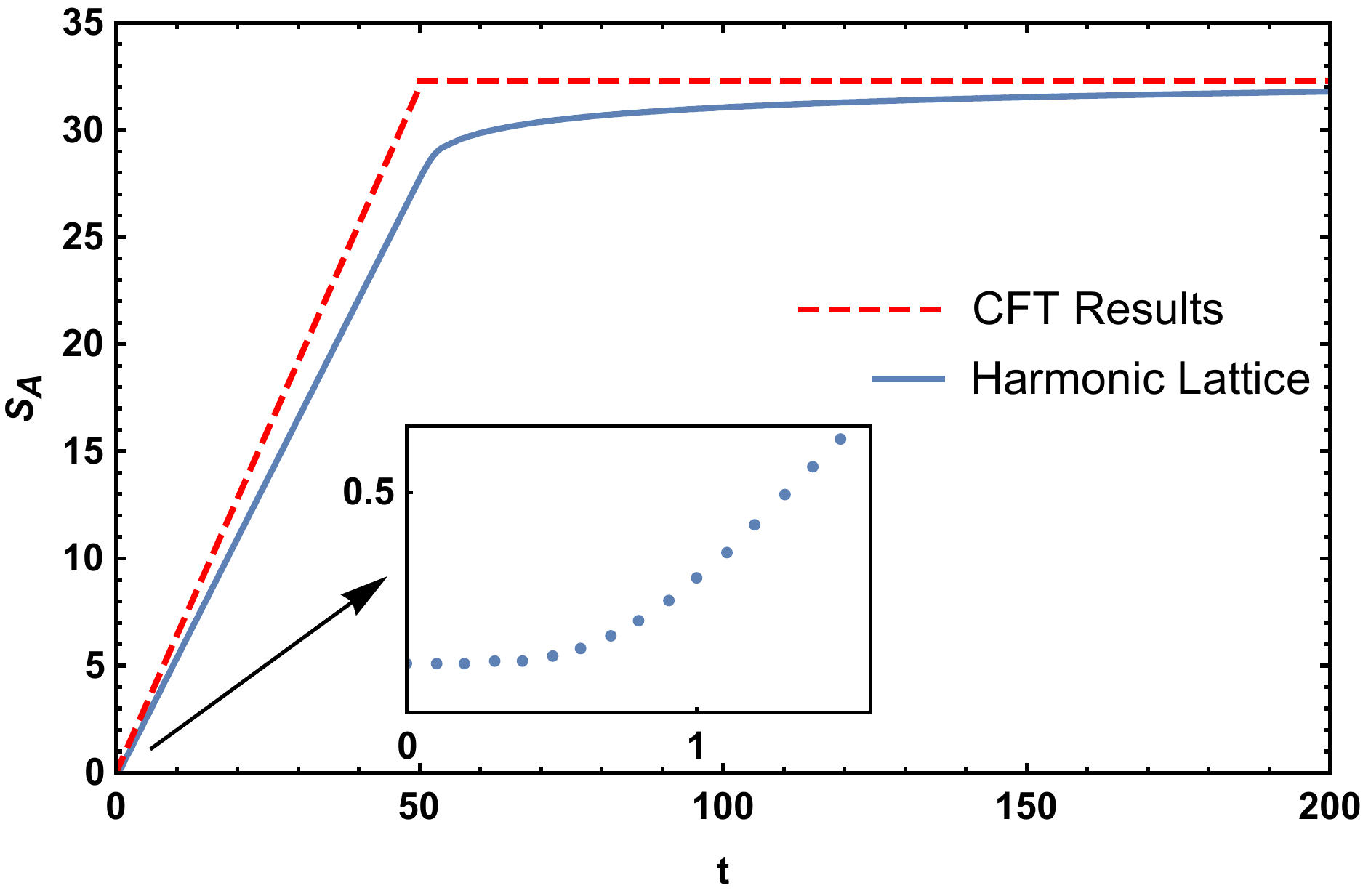}
\hspace*{0.5cm}
\includegraphics[scale=0.4]{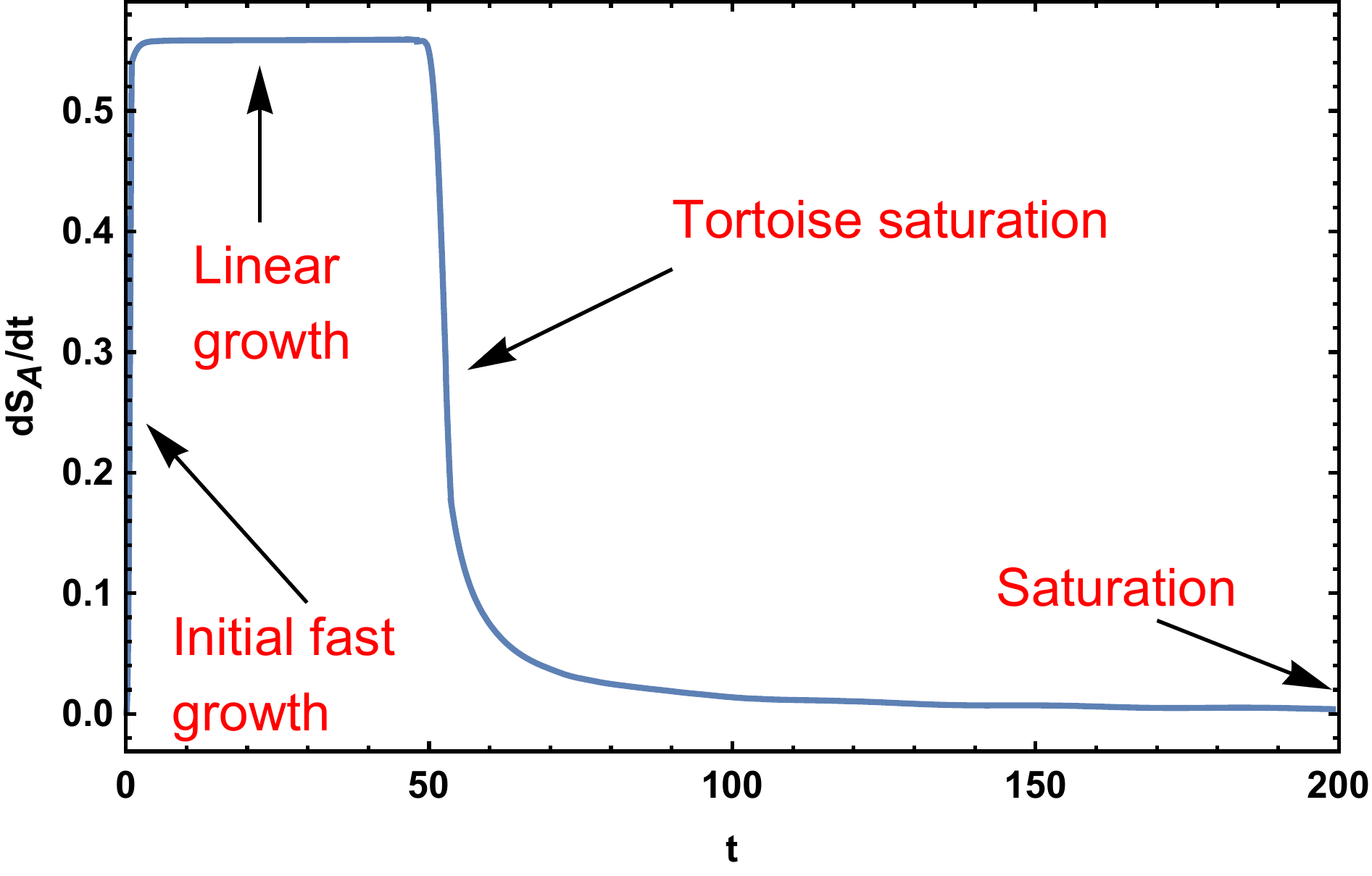}
\end{center}
\caption{Numerical data showing entanglement entropy as a function of time in a theory with relativistic scaling. \textit{Left:} CFT prediction vs. harmonic lattice simulation with periodic BC. The numerical results correspond to $\ell=100$ and $m_0=1,\;m=10^{-6}$. \textit{Right:} Time derivative of EE during the thermalization process.}
\label{fig:ccresults}
\end{figure}

In figure \ref{fig:ccresults} we have summarized mainly the results of \cite{Calabrese:2005in}, where the numerical results are obtained using a harmonic lattice model.    
Note that although in the case of a CFT the saturation happens instantaneously, lattice simulation shows a mild transition. In order to clarify different scaling regimes of the process, in the right panel we have also numerically plotted the time derivative of EE. At the very beginning there is quadratic growth regime\footnote{Although this behaviour was first found in the holographic context \cite{Liu:2013iza}, but it can also be captured by 2d CFT techniques similar to \cite{Calabrese:2005in}, see for instance section 2.3.1. of \cite{Jarvela:2014pbd}.}$^,$\footnote{The quadratic growth regime is before local-equilibrium and is not captured by the quasi-particle picture.} after which there is the well-known linear growth. The linear growth is followed by an extremely slower growth which is argued in \cite{Cotler:2016acd} to logarithmically depend on time.\footnote{An analytical argument in support of $\log t$ behaviour is recently given in \cite{Chapman:2018hou}.} We will denote this regime by \textit{tortoise saturation} in what follows and will also derive an analytic time dependence for this regime in a certain limit of mass quenches. Due to the mild transition, i.e. the existence of the tortoise saturation regime, saturation time is postponed \cite{Calabrese:2005in,Cotler:2016acd}. We will elaborate on this point both in Lorentzian and in Lifshitz theories.

There are technical problems comparing between harmonic lattice and CFT results. Utilizing the harmonic lattice model, the straightforward choices is to impose periodic boundary condition which leads to a translational invariant model, although a discrete one. But this model in general suffers from IR divergence due to the existence of a zero mode. This zero mode and a pile of other slow modes which come into the game as soon as the model needs an IR cut-off are actually responsible for this mild transition. On the other hand one can impose a Dirichlet boundary condition on both ends of the lattice and sticking the entangling region to one of them in order to get around the problem. Although this makes the transition sharper since the zero mode is killed, but does not solve the problem and the price is loosing translational invariance.

The qualitative behavior of $S_A(t)$ during this relaxation process is well-known to be understood by the quasi-particle picture\cite{Calabrese:2005in}. In this picture the entanglement between a subregion $A$ and its complement $\bar{A}$ is carried by a uniform density of free streaming quasi-particles. A pair of entangled quasi-particles is created at each spatial point and the entropy between $A$ and $\bar{A}$ is measured by the number of quasi-particles which a pair is in $A$ and the other pair is in $\bar{A}$. As a consequence of causality the propagation of quasi-particles constrained to be inside the light-cone such that the massless quasi-particles move along the null rays. Using this simple scenario it has been shown that the saturation time is given by $t_s=\ell/(2v_g)$, where $v_g$ denotes the group velocity of entanglement propagation quasi-particles which in a critical theory is $v_g=1$.

\begin{figure}
\begin{center}
\includegraphics[scale=0.35]{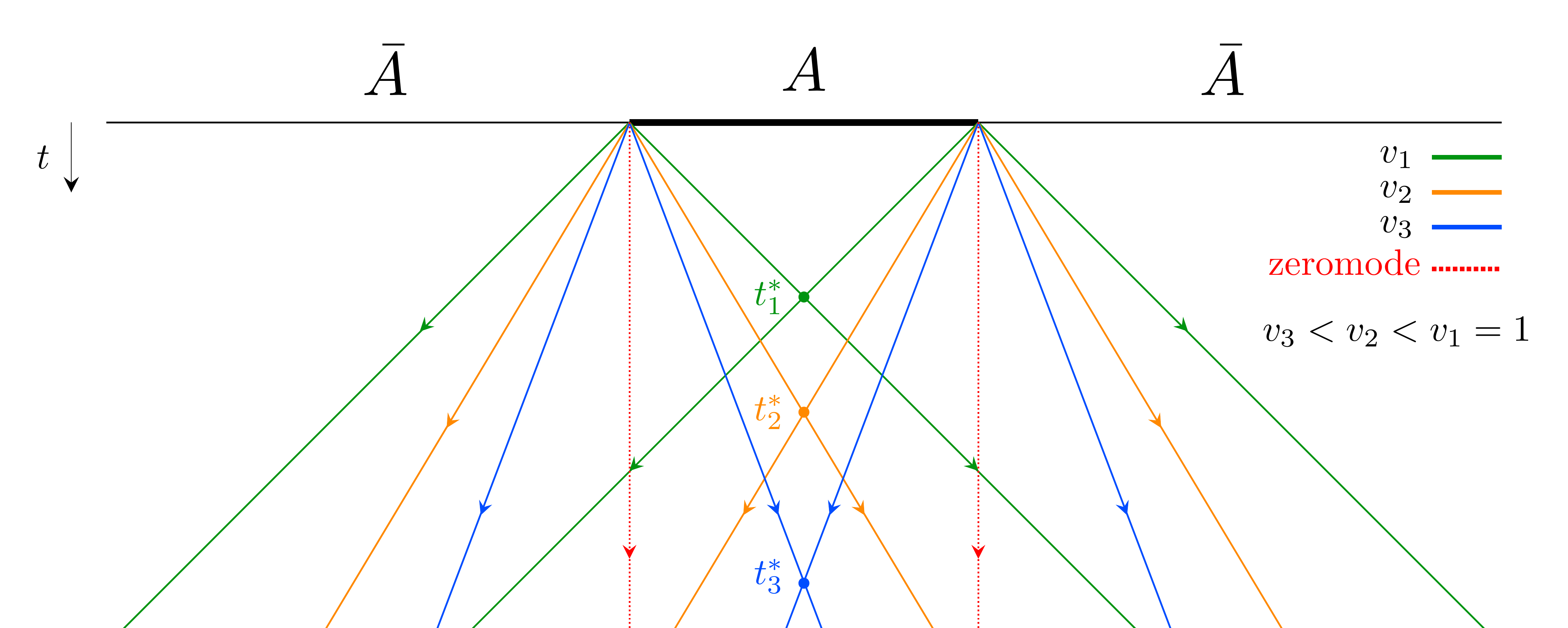}
\end{center}
\caption{Alba-Calabrese quasi-particle picture for a relativistic free boson and three distinct modes together with the zero mode. The saturation time for each mode is shown with $\{t^*_1,t^*_2,t^*_3\}$.}
\label{fig:qp0}
\end{figure}

Recently an analytic description for the time evolution of EE after a quantum quench based on the quasi-particle picture was proposed in \cite{Alba:2017lpnas, Alba:2017lvc}.\footnote{It has been also extended to multipartite subregions in \cite{Alba:2018hie}, to more general initial states in \cite{Alba:2017qag, Bastianello:2018fvl, Bertini:2018kpy, Bertini:2018ymp} and to study Renyi entropies in \cite{Mestyan:2018fwt, Alba:2017obx, Alba:2017kdq}.} It was shown that considering a quasi-particle picture for spread of entanglement and also knowing the late time stationary state provided by the Bethe ansatz, one can find an analytic description for time evolution of EE. This construction works for several integrable models including non-interacting bosonic and fermionic systems. Since we will utilize this picture to understand our results, we will shortly review its main features in the following.  

According to this description the general prediction for time dependence of EE in a $2d$ theory for a subregion with length $\ell$ is given by \cite{Alba:2017lvc}
\bea\label{analytic}
S(t)=2t\int_{2|v_g|t<\ell}dk |v_g|s(k)+\ell\int_{2|v_g|t>\ell}dk s(k),
\eea
where $s(k)$ is the entropy density and $v_g\equiv d\omega_k/dk$ is the group velocity of the corresponding quasi-particles with momentum $k$.

To have a better intuitive understanding of what Eq.\eqref{analytic} means, we have illustrated the physics in figure \ref{fig:qp0}. In general there is an infinite number of modes in the model and in this figure we have shown four of them to describe the physics in a relativistic model. The fastest mode is shown in green and two slower ones in orange and blue together with the zero mode in red. When a global quench happens at any spatial point, corresponding to each mode a pair of quasi-particle starts to move around back to back. Different quasi-particles have different group velocities bounded by $|v_g|\le1$. Each mode has a saturation time with regard to its group velocity given by $\ell/(2v_g)$. The saturation time of each mode (for a single interval) is by definition the point where the corresponding rays from the two ends of the region intersect. These are shown by $\{t^*_1,t^*_2,t^*_3\}$. Obviously since $v_{\text{zero mode}}(=0)<v_3<v_2<v_1$, one would expect $t^*_{\text{zero mode}}>t^*_3>t^*_2>t^*_1$. The zero mode saturation time is infinite. Due to this intuitive picture one can understand the physics of Eq.\eqref{analytic}. At any time $t$, there are a number of modes which are fast enough to be saturated some time before $t$, these modes no more contribute to the time evolution of EE and form the second term in Eq.\eqref{analytic}. At the same time the slower modes which satisfy $\ell/(2v_g)>t$, still contribute to the time evolution and form the first term in Eq.\eqref{analytic}. The smaller $t$ is, the larger the number of modes contributing to the first term.  
 
\begin{figure}
\begin{center}
\includegraphics[scale=0.4]{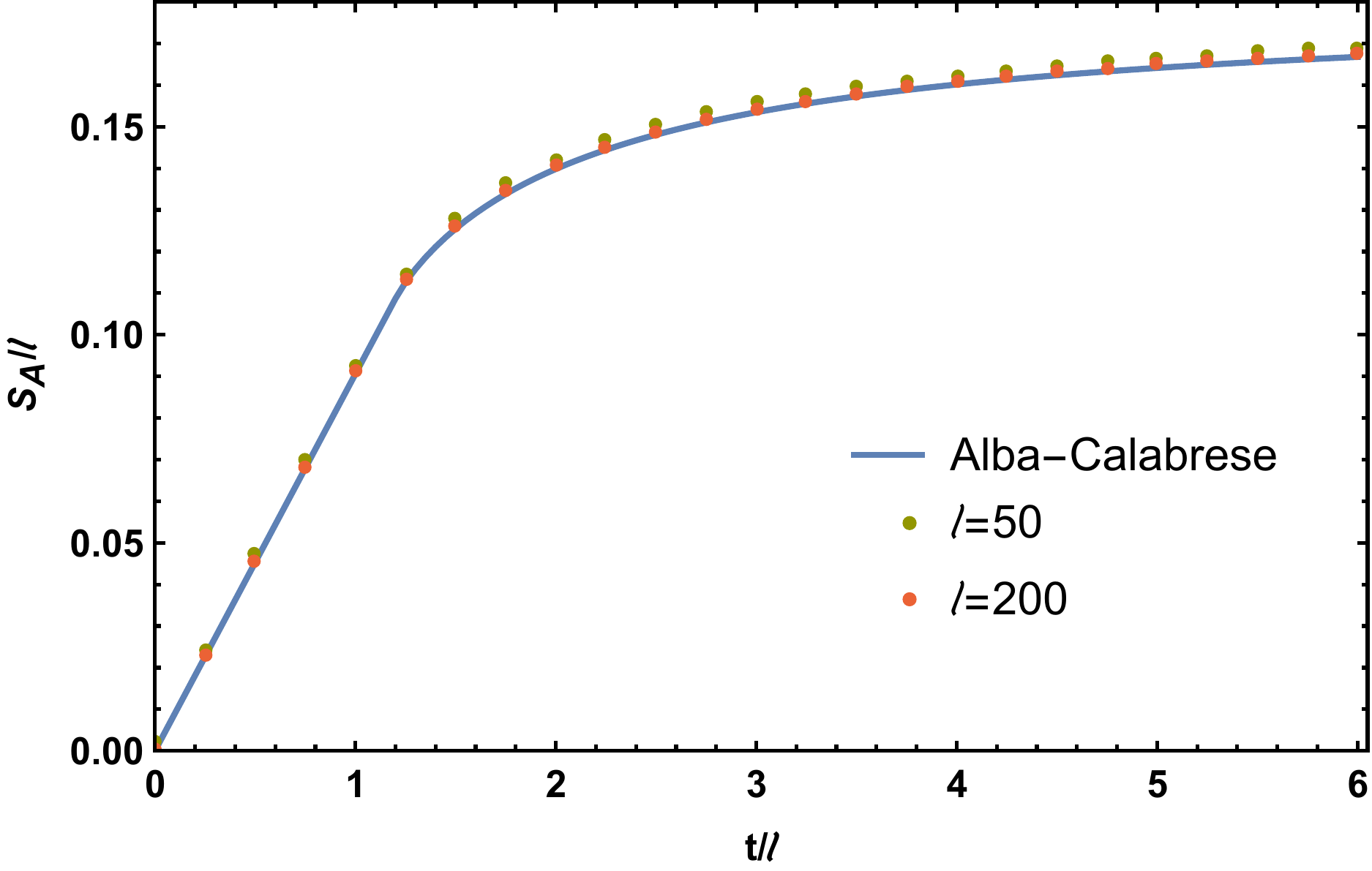}
\hspace{5mm}
\includegraphics[scale=0.4]{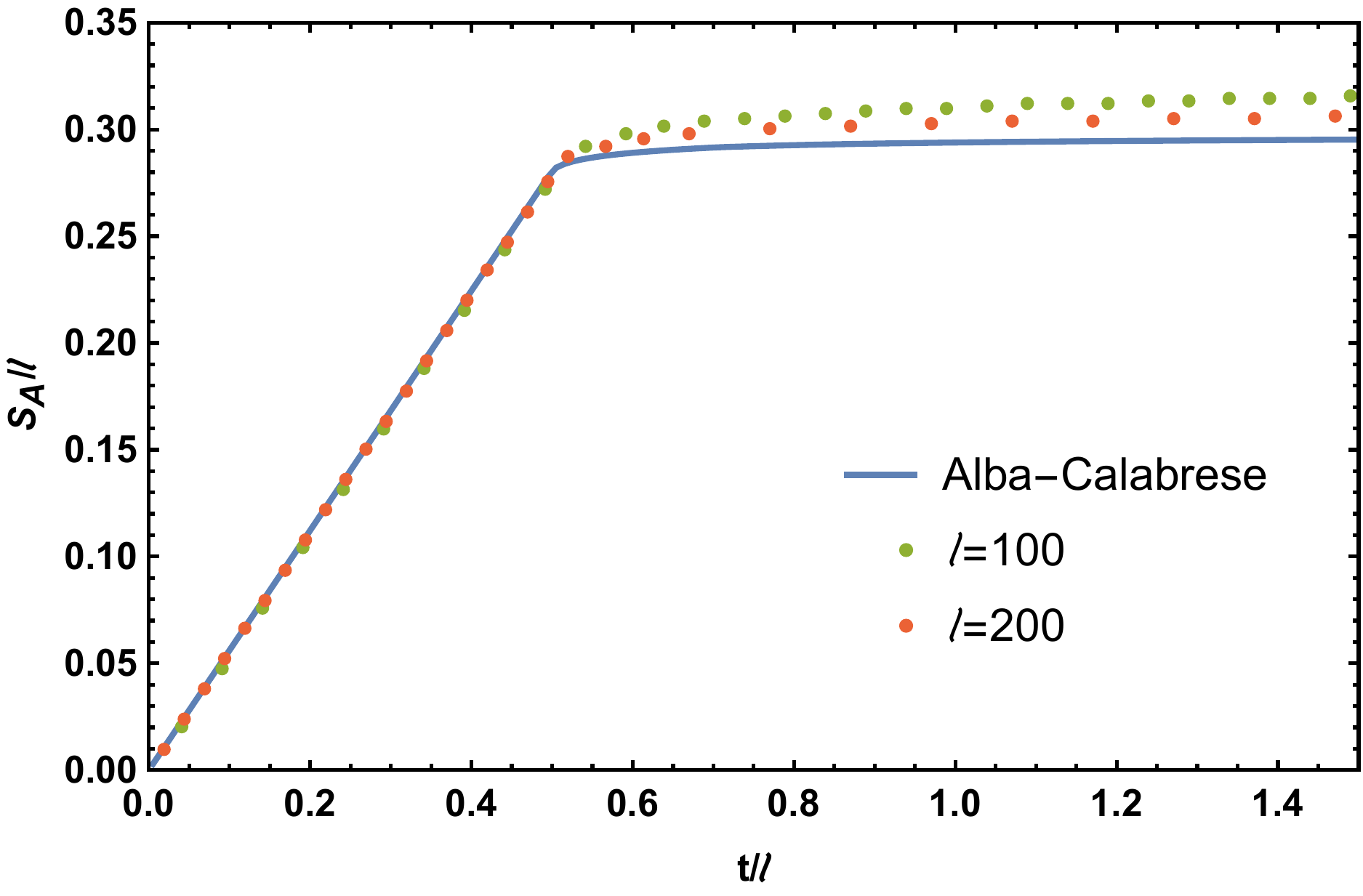}
\end{center}
\caption{Analytic versus numerical results for the evolution of EE in harmonic lattice. In the left panel we have set $m_0=1$ and $m=2$ and in the right panel we have set $m_0=1$ and $m=0$. In the right panel the effect of the zero mode which is not captured by the analytical picture causes small deviation for $t>\ell/2$.}
\label{fig:alba}
\end{figure}

Indeed in \cite{Alba:2017lpnas} it was shown that for certain integrable models, $s(k)$ can be fixed employing the equivalence between the entanglement and the thermodynamic entropy in the stationary state. Note that the thermodynamic entropy can be calculated from the generalized Gibbs ensemble describing the stationary state in terms of the expectation value of mode occupation number as follows
\bea
s(k)=\frac{1}{2\pi}\left((n_k+1)\log (n_k+1)-n_k\log n_k\right),
\eea
where $n_k=\langle\hat{n}_k\rangle_{\rm GGE}$. Also note that because $\hat{n}_k$ is an integral of motion, $n_k$ can be found using the initial state, i.e.,
\bea
\langle\hat{n}_k\rangle_{\rm GGE}=\langle a_k^{\dagger}a_k\rangle_{\rm GGE}=\langle \psi_0|a_k^{\dagger}a_k|\psi_0\rangle.
\eea
In the case of a free scalar theory one finds \cite{Calabrese:2007rg}
\bea\label{nk}
n_k=\frac{1}{4}\left(\frac{\omega_k}{\omega_{0,k}}+\frac{\omega_{0,k}}{\omega_k}\right)-\frac{1}{2},
\eea
where $\omega_{0,k}$ ($\omega_{k}$) is the frequency before (after) the quantum quench. Now we are equipped with all we need to calculate the time dependence of EE using Eq.\eqref{analytic}. In figure \ref{fig:alba} we demonstrate the evolution of EE predicted by Eq.\eqref{analytic} and compare it with the numerical results for different values of $\ell$ once with $m_0=1, m=2$ and once with $m_0=1, m=0$. Note that as we increase the length of the entangling region, we find better agreement. For the first family where $m\neq 0$ the results for 200 number of sites is almost the same as the thermodynamics limit but for massless post-quench Hamiltonian it does not coincide with the quasi-particle picture which we believe is due to the zero mode effect. We will try to analyse this point in details in what follows.

Remarkably, there exists an interesting relation between the spectrum of the quasi-particles and the saturation time of the EE. As the concluding note for this section we explain this relation in the simplest case, i.e., the harmonic lattice. In a 2$d$ QFT with relativistic scaling the lattice dispersion relation for the corresponding quasi-particles is given by Eq.\eqref{dispersion} (setting $z=1$) as follows\footnote{Note that in the following without loss of generality we consider the continuum limit of the dispersion relation.}
\bea\label{eq:disz1}
\omega_k=\sqrt{m^{2}+\left(2\sin \frac{k}{2}\right)^{2}}.
\eea
The group velocity of the quasiparticles from the above equation is
\bea\label{vg}
v_g(k)\equiv \frac{d\omega_k}{dk}=\frac{\sin k}{\sqrt{m^{2}+\left(2\sin \frac{k}{2}\right)^{2}}}.
\eea
Now one can find the mode with the maximum group velocity as follows
\bea\label{vgmax}
\frac{dv_g}{dk}\bigg|_{k=k_{\text{max}}}=0\;\;\;\;\;\Rightarrow\;\;\;\;\;\kappa^2+m^2\kappa-m^2=0,
\eea
where $\kappa=2\sin^2 \left(\frac{k_{\text{max}}}{2}\right)$.
Regarding Eq.\eqref{vg} and Eq.\eqref{vgmax} the following comments are in order 
\begin{itemize}
\item 
To trace the effect of tortoise modes (those with small momenta) more precisely, we look at the $k\to0$ limit of the group velocity. The behaviour is as follows
$$v_g(k\to0)\approx \frac{k}{m}+\mathcal{O}\left(k^2\right).$$ 
One should note that since we are considering a translational invariant lattice, which the dispersion relation is given in Eq.\eqref{eq:disz1}, there is an IR regulator in the model. Thus to track the effect of the zero mode in a scale invariant post-quench state one should always be careful about the order of the $m\to0$ and $k\to0$ limits. Because of the existence of the IR regulator, one should take the $k\to0$ limit first. Note that $k=0$ is a permissible mode in this model thus whatever the IR regulator (the mass of the post-quench state) is, the zero mode has vanishing group velocity and modes with small momenta ($k\ll m$) also have very small group velocities. These modes are responsible for tortoise saturation after all other fast modes are saturated.    

\item For a massive quasiparticle we have $\kappa=-\frac{m^2}{2}+m\sqrt{1+\left(\frac{m}{2}\right)^2}$ and the maximum group velocity is given by
\bea
v_g^{\text{max}}=\sqrt{1+\frac{m^2}{2}-m\sqrt{1+\left(\frac{m}{2}\right)^2}},
\eea
which shows that the maximum velocity keeps decreasing as a function of the mass parameter.

\item
From the above picture in principle it is straightforward to workout the analytical behaviour of EE after saturation time. Here there are tortoise modes which do not saturate at finite time, but one can still define an effective saturation time regarding to the saturation of the fastest mode $t_s^{\text{max}}$. In general there is a maximum momentum which we denote by $k_\text{max}$ corresponding to each mode. In principle the time dependent part of entropy 
\be\label{eq:Sanalytic}
\mathcal{S}(t)=\int_{|k|<k_\text{max}(t)}dk\,s(k)
\ee
gives the time dependence of entropy after $t_s^{\text{max}}$. Even for the free bosonic case this is not analytically computable but in a certain limit which one considers a quench from a very heavy state ($m\ell\gg1$) to a gapless state  ($m\ell\ll1$) the above integral can be performed analytically which leads to (for $t>\ell/2$)
\be
\mathcal{S}(t)=\mathcal{S}_0+\frac{1}{\sqrt{t^2-\ell^2/4}}\left(c_1+c_2\log{\sqrt{t^2-\ell^2/4}}\right)m+\mathcal{O}\left(m^2\right) 
\ee
where $\mathcal{S}_0$ is the value of EE at $t=\ell/2$, $c_1$ and $c_2$ are constants depending on $m_0$, $m$ and $\ell$. The key point to work out this behaviour is that the above intuitive picture lets us to think about the time dependence of EE as the time dependence of $k_\text{max}$.

It is worth to mention that the above behaviour, although is found for a quench from highly gapped to a gapless Hamiltonian, but numerical data show that it works well even in the regime of our study where $m_0\sim \mathcal{O}(1)$ and $m=10^{-6}$. 
\item Thinking for a while about the continuum scalar field theory, the qualitative features of the previous results do not change. In this case the group velocity can be obtained from the continuous dispersion relation $\omega=\sqrt{m^2+k^2}$ as follows
\bea\label{vgconti}
v_g(k)=\frac{k}{\sqrt{m^2+k^2}}.
\eea
Once again the existence of massive particles with vanishingly small momentum (tortoise modes), leads to a tortoise saturation regime for EE at late times.

\item
To get around these tortoise modes, one approach is to consider mode-dependent mass quench, i.e., $m(k)$\cite{Cotler:2016acd}. In this case the group velocity becomes
\bea
v_g(k)=\frac{k+m(k)m'(k)}{\sqrt{m(k)^2+k^2}},
\eea
where we should impose $v_g(k\sim 0)\rightarrow v_0(>0)$ to prevent the excitation of tortoise modes. Any mass function with the specific behavior of $m(k\sim 0)=m_0+ v_0 k+\cdots$ satisfies this condition and removes these modes from the spectrum of quasi-particles\footnote{Note that in order to have a finite injected energy during the quench scenario we should impose $m(k\sim \infty) \rightarrow 0$.}. Considering these family of quenches, one no longer finds a finite saturation time with no tortoise saturation regime. It is important to note that according to Eq.\eqref{vgconti} the massless quasi-particles with linear dispersion relation ($\omega\sim k$) move along the null rays with the maximum momentum independent velocity, $v_g(k)=1$.
\end{itemize}

\section{Entanglement Propagation in Lifshitz Theories}\label{sec:EEz}
In this section we study how EE propagates in theories with Lifshitz scaling. To be more precise, we study post-quench states both with $m=0$ and $m>0$. We first study analytically the quasi-particle picture for Lifshitz theories, modelled by Lifshitz harmonic lattice, after which we present numerical studies of EE and discuss about the physics of propagation of entanglement in different scaling regimes.

\subsection{Analytic Description}\label{subsec:analytic}
In principle since our Lifshitz field theory of interest is a generalization of Klein-Gordon theory but the spatial correlations are stretched out via the dynamical exponent, one would naturally expect a similar analysis to what we have reviewed in the previous section from  \cite{Alba:2017lpnas,Alba:2017lvc} is valid in this theory.

\begin{figure}
\begin{center}
\includegraphics[scale=0.4]{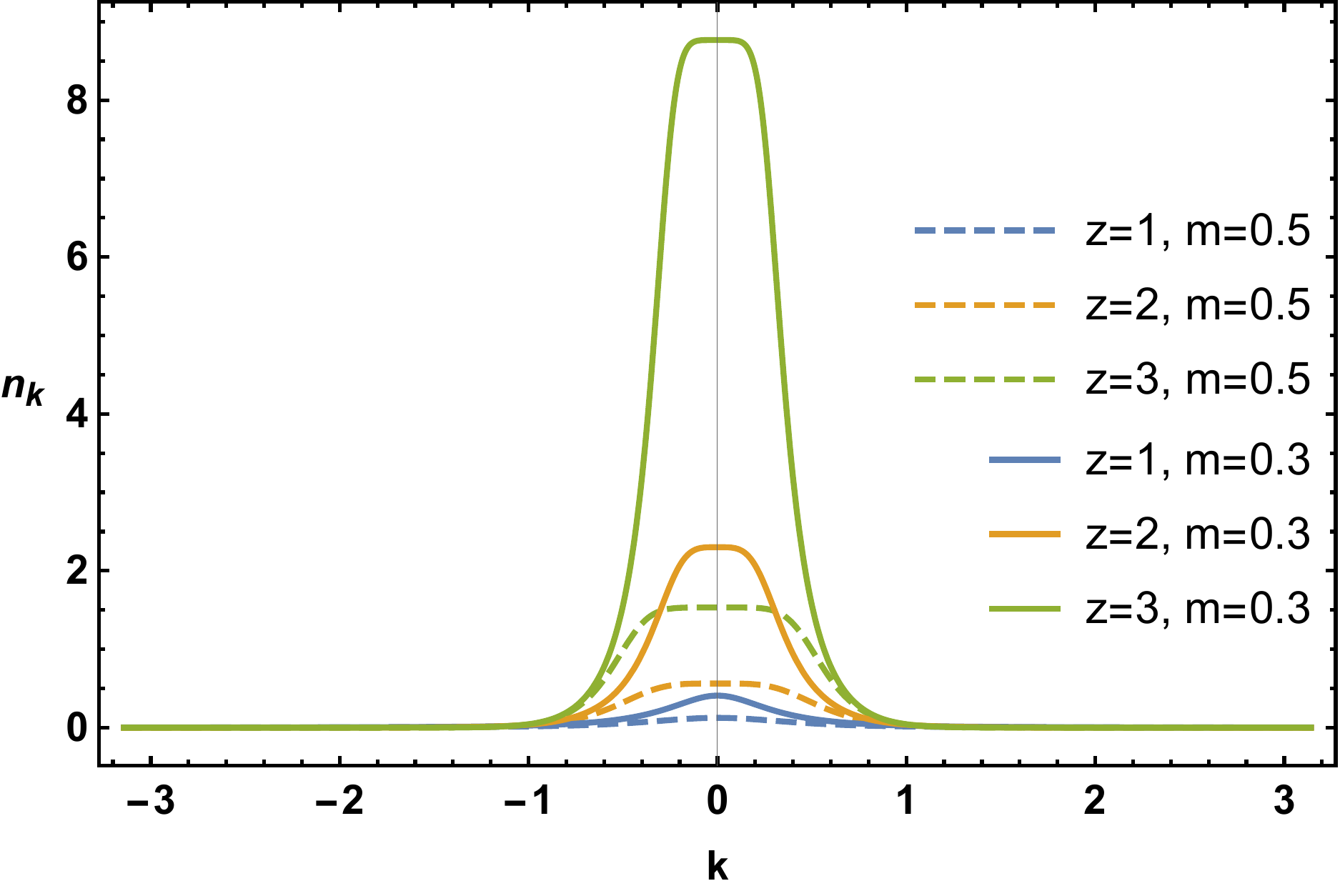}
\hspace{0.5cm}
\includegraphics[scale=0.4]{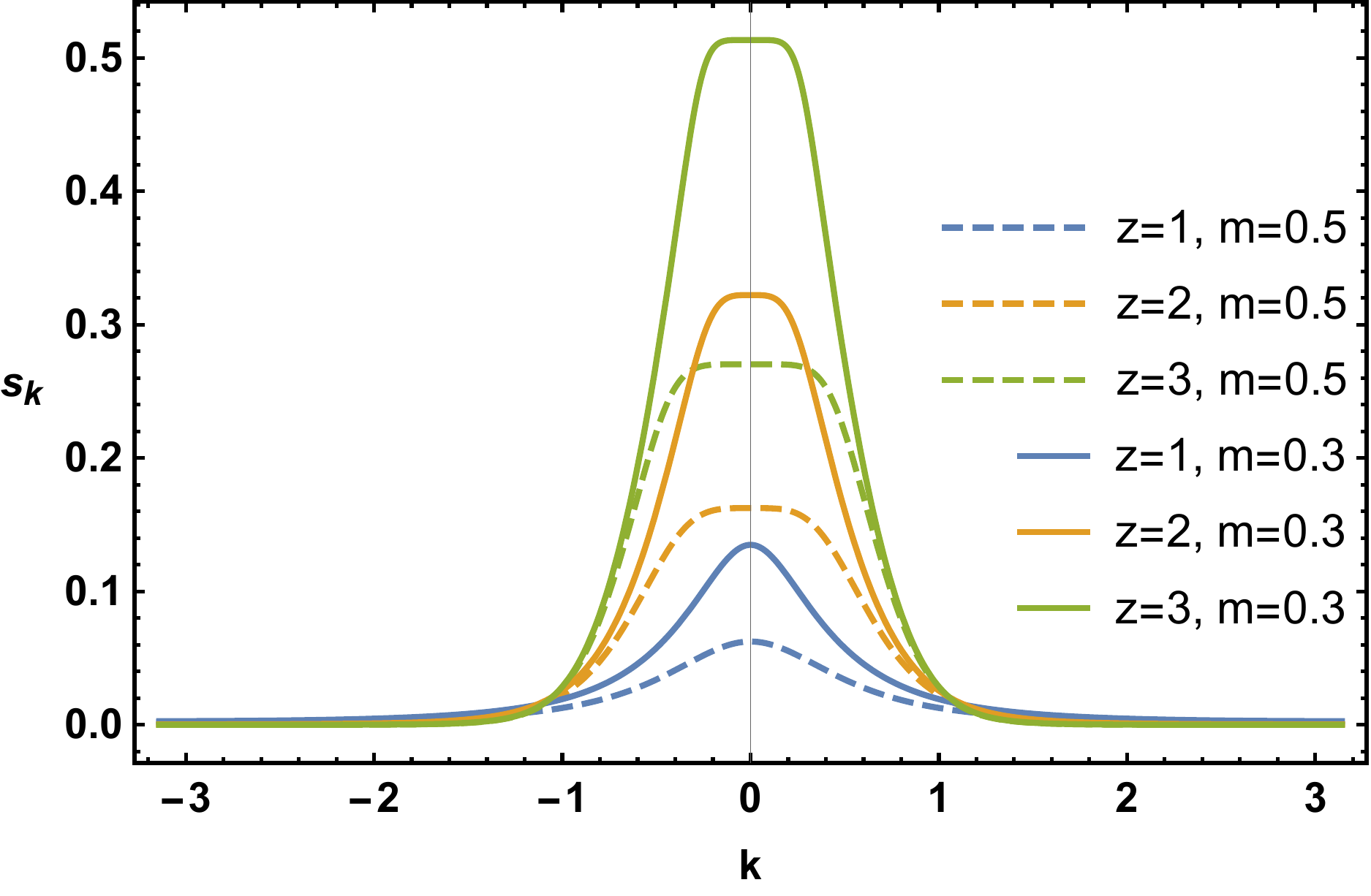}
\end{center}
\caption{Mode occupation number and entropy density as a function of $k$ for different values of the dynamical exponent. Here we set $m_0=1$. We have shown two values of post-quench mass to see how fast the $n(k)$ and $s(k)$ diverge for small values of mass. Comparing these two values of $m$ gives a sense of how fast the occupation number (and entropy density) diverge in the massless limit. The same behaviour is correct for fixed post-quench mass and small pre-quench mass.}
\label{fig:nksk}
\end{figure}

The intuitive description we discussed in the previous section, simply shows that similar to the relativistic case, the general time dependence of EE predicted in a $2d$ free scalar theory with Lifshitz sacling is given by Eq.\eqref{analytic}. So what we need is to workout the exact dependence of $n(k)$, $s(k)$ and $v_g$ on the dynamical exponent. Using the dispersion relation Eq.\eqref{dispersion} a strightforward computation gives
\bea\label{vgz}
v_g(z)=z\frac{\left(2\sin \frac{k}{2}\right)^{2z-2}\sin k}{\sqrt{m^{2z}+\left(2\sin \frac{k}{2}\right)^{2z}}}.
\eea
Also similar to the relativistic case the entropy density in terms of the expectation value of mode occupation numbers in the initial state is given by Eq.\eqref{nk} where we should consider the dependence of $\omega_k$ and $\omega_{0,k}$ on $z$ as follows
\bea\label{Omega}
\omega_k=\sqrt{m^{2z}+\left(2\sin \frac{k}{2}\right)^{2z}},\;\;\;\;\;\omega_{0,k}=\sqrt{m_0^{2z}+\left(2\sin \frac{k}{2}\right)^{2z}},
\eea
where $m_0$ and $m$ are the mass parameters before and after the quantum quench respectively. In general the value of $n(k)$ and $s(k)$ increase at a given momentum as we increase the dynamical exponent. In other words the contribution of individual quasi-particles to the thermodynamic entropy of the generalized Gibbs ensemble describing the steady state increases due to a $z>1$. In the following we will present some results which the role of tortoise modes becomes extremely important. As an example we have plotted the occupation number and the entropy density in figure \ref{fig:nksk} for quenches from a fixed pre-quench state ($m_0=1$) to different values of smaller post-quench mass parameters. We have considered these parameters to figure out what happens toward quenching to gapless models ($m\to0$). One can easily check from the expression of $n(k=0)$ which is 
$$n(k=0)=\frac{1}{4}\left(\left(\frac{m}{m_0}\right)^z+\left(\frac{m_0}{m}\right)^z\right)-\frac{1}{2},$$
that the occupation number (and also the entropy density) diverges in the following three cases: $m_0/m\ll1$, $m/m_0\ll1$, and $z\gg1$. In the case of scale invariant system in principle the post-quench mass vanishes. But here due to the IR cut-off, this case is a special case of $m/m_0\ll1$. We will show in what follows that the numerical results deviate from the quasi-particle picture in these three regimes, although the picture works perfectly in other regimes. 

In the following sections after describing the quasi-particle picture we will present numerical results where our main focus is comparing the vanishing and non-vanishing mass parameters in the post-quench state to inquire our understanding of the quasi-particle picture.

\subsection{Quasi-particle Picture}
As we discussed in section \ref{sec:EEz1}, the spectrum of quasi-particles together with the notion of causality interestingly describes the linear growth and tortoise saturation regimes. Here we explore the role of $z$ in this scenario focusing on 2$d$ theories in order to avoid any complication arising in case of more than one velocity component. In general we would expect the qualitative picture should be straightforwardly generalized to higher dimensions. Although the notion of causality is a totally subtle notion in these theories, we show that (except in certain cases which the tortoise mode contributions become dominant) different scaling regimes during the relaxation of the system can be perfectly described by Alba-Calabrese quasi-particle picture in presence of $z>1$ dynamical exponents. Comparing to the $z=1$ case there are interesting features in $z>1$ which we will describe in the following. 

\begin{figure}
\begin{center}
\includegraphics[scale=0.4]{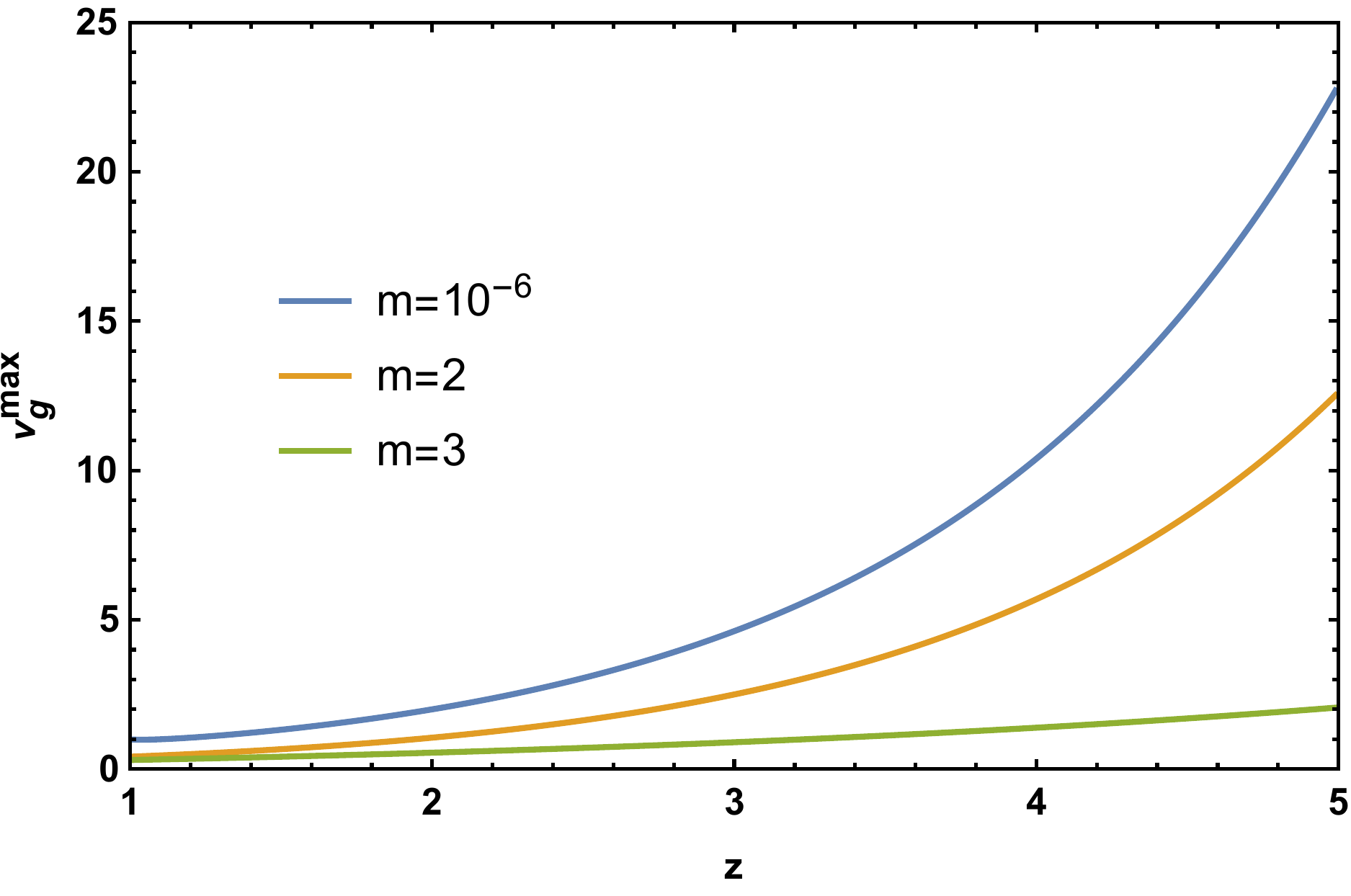}
\hspace*{0.5cm}
\includegraphics[scale=0.4]{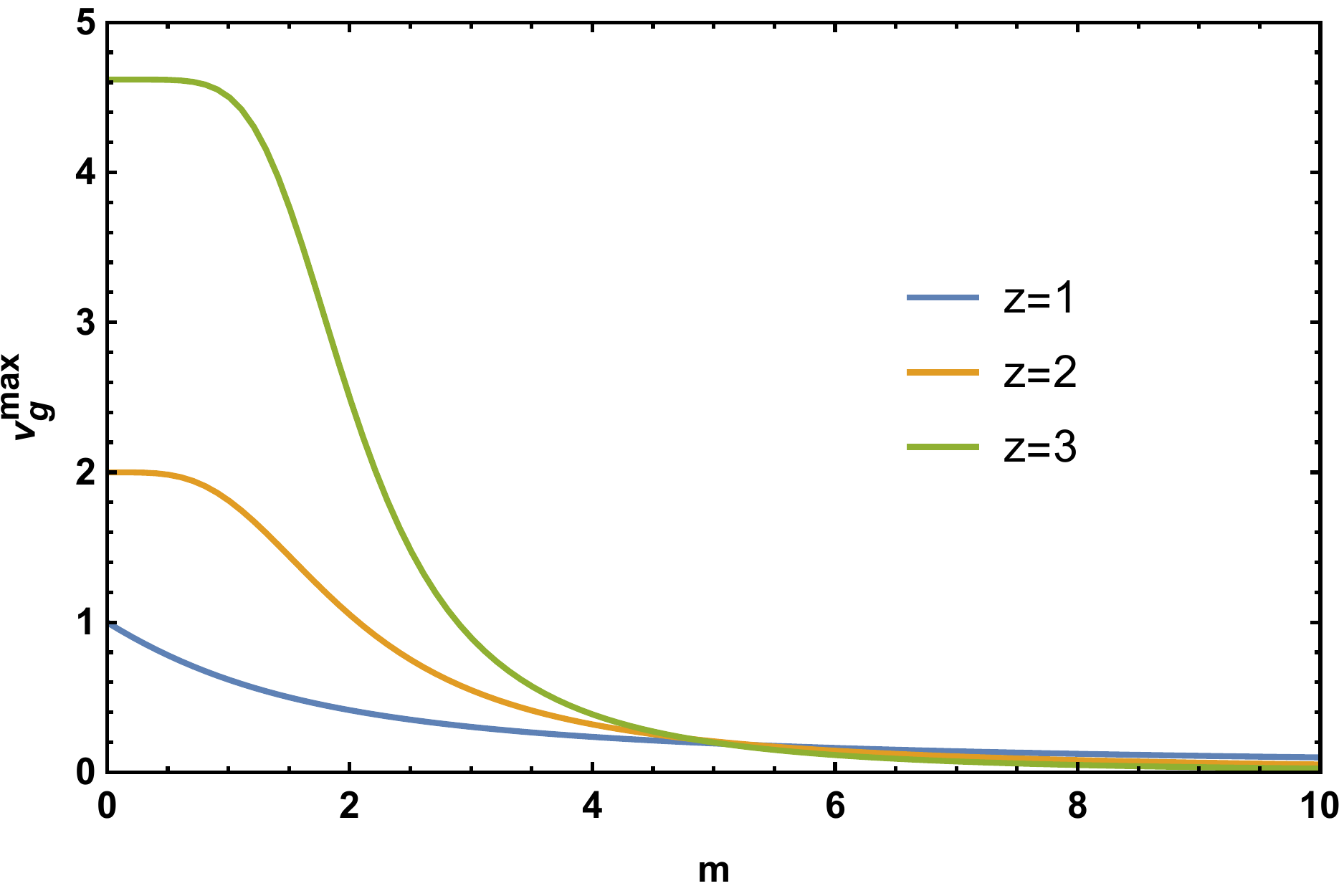}
\end{center}
\caption{Left: $v_g^{\rm max}$ as a function of $z$ for different mass parameters. Right: $v_g^{\rm max}$ as a function of $m$ for different values of the dynamical exponent.}
\label{fig:vmaxz}
\end{figure}

The group velocity for quasi-particles in a Lifshitz harmonic lattice is given by Eq.\eqref{vgz}. Using this relation
\bea
v_g(z)=2^{2z-1}z\frac{\cos\frac{k}{2}\,\sin^{2z-1} \frac{k}{2}}{\sqrt{m^{2z}+\left(2\sin \frac{k}{2}\right)^{2z}}},
\eea
the maximum group velocity can be derived by first solving $dv_g/dk=0$ for the maximum momenta. This can be done analytically for small post-quench mass (for $z>1$) parameter as
$$k_{\text{max}}=\cos^{-1}\left(\frac{2-z}{z}\right)+\left(\frac{z}{4}\right)^z\left(z-1\right)^{-z-\frac{1}{2}}m^{2z}+\mathcal{O}\left(m^{4z}\right)$$
which gives
\bea\label{vgm0max}
v_g^{\rm max}(z)=2^{z-1} \sqrt{z}\left(\frac{z-1}{z}\right)^{\frac{z-1}{2}}-2^{-z-2}\sqrt{z}\left(\frac{z-1}{z}\right)^{-\frac{z+1}{2}}m^{2z}+\mathcal{O}\left(m^{4z}\right).
\eea
According to the above result a few comments are in order:

\begin{itemize}
\item For $z=1$ in the massless limit we have $v_g^{\rm max}=1$ which is consistent with the group velocity of quasi-particles which we reviewed in section \ref{sec:EEz1}. Also note that the mass correction is negative independent of $z$ which shows that as expected intuitively, the velocity is a decreasing function of the mass parameter. 

\item
Similar to what we discussed in Eq.\eqref{eq:Sanalytic}, here also it is possible to extract an analytical temporal behaviour of EE for $z>1$. The most straightforward case is $z=2$ which the same analysis again in a limit which a Hamiltonian with a large mass is quenched to a tiny mass Hamiltonian one finds 
\be
\mathcal{S}(t)=\mathcal{S}_0+\frac{1}{t}\left(c_1+c_2\log{t}\right)m+\mathcal{O}\left(m^2\right),
\ee
where $\mathcal{S}_0$ is the value of EE at $t=\ell/(2v_g^{\text{max}})$ and $v_g^{\text{max}}$ is given by Eq.\eqref{eq:vgmaxpert} for $z=2$, $c_1$ and $c_2$ are constants depending on $m_0$, $m$ and $\ell$.
The analysis is a bit hard to be extended for higher values of $z$ because of a technical problem which one cannot easily solve for the generic time dependence of $k_{\text{max}}(t)$.

\item For $z\to\infty$ where the corresponding scalar theory given by Eq.\eqref{action} becomes strongly non-local the maximum group velocity diverges, i.e., 
\bea\label{eq:vgmaxpert}
v_g^{\rm max}\big|_{z\rightarrow \infty}= 2^{z-1}\sqrt{\frac{z}{e}}+\mathcal{O}\left(\frac{1}{z}\right).
\eea
Note that the existence of a maximum group velocity in this non-relativistic model, Eq.\eqref{vgm0max} is consistent with the general expectation of the existence of Lieb-Robinson bound in local (laticized) QFTs \cite{LR, LRB1, LRB2, toAppear1}\footnote{The existence of Lieb-Robinson bounds in discrete systems but with infinite dimensional Hilbert space at each site, such as Harmonic lattice model, has been proved in \cite{LRB1, LRB2}. The same proof works for Lifshitz harmonic model replacing the corresponding dispersion relation. We would like to emphasize that these bounds are not strong enough to lead to a physically reasonable Lieb-Robinson velocity.}.\footnote{For a related study in non-relativistic theories see \cite{Roberts:2016wdl}.} For any finite $z$ there is an upper bound on the propagation velocity of the quasi-particles. The above relation shows that in the non-local limit where the Lieb-Robinson bound is expected to break down, there is no upper bound on velocity (see section \ref{sec:LatvCon} and \cite{toAppear1}).

\begin{figure}
\begin{center}
\includegraphics[scale=0.28]{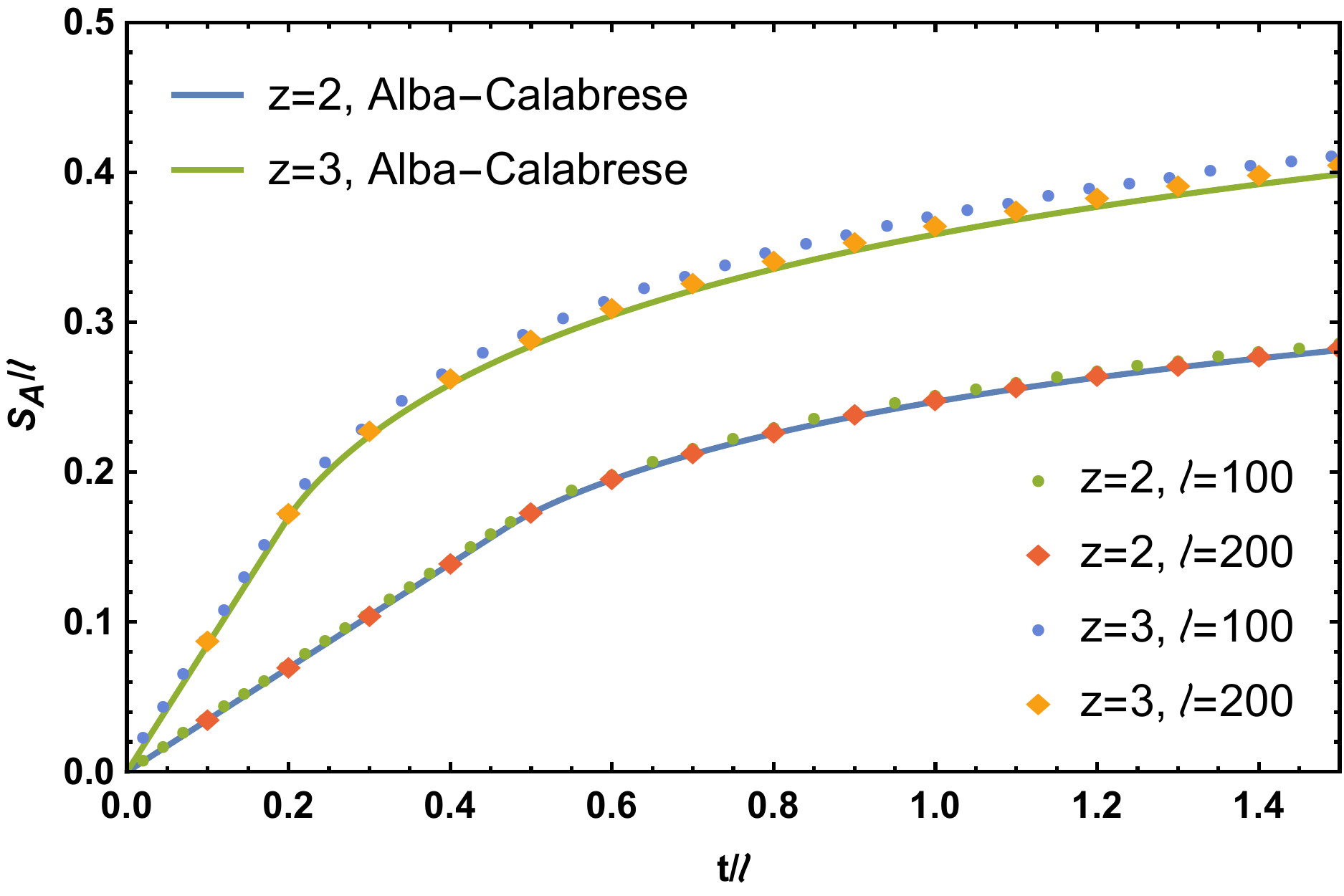}
\hspace{1mm}
\includegraphics[scale=0.28]{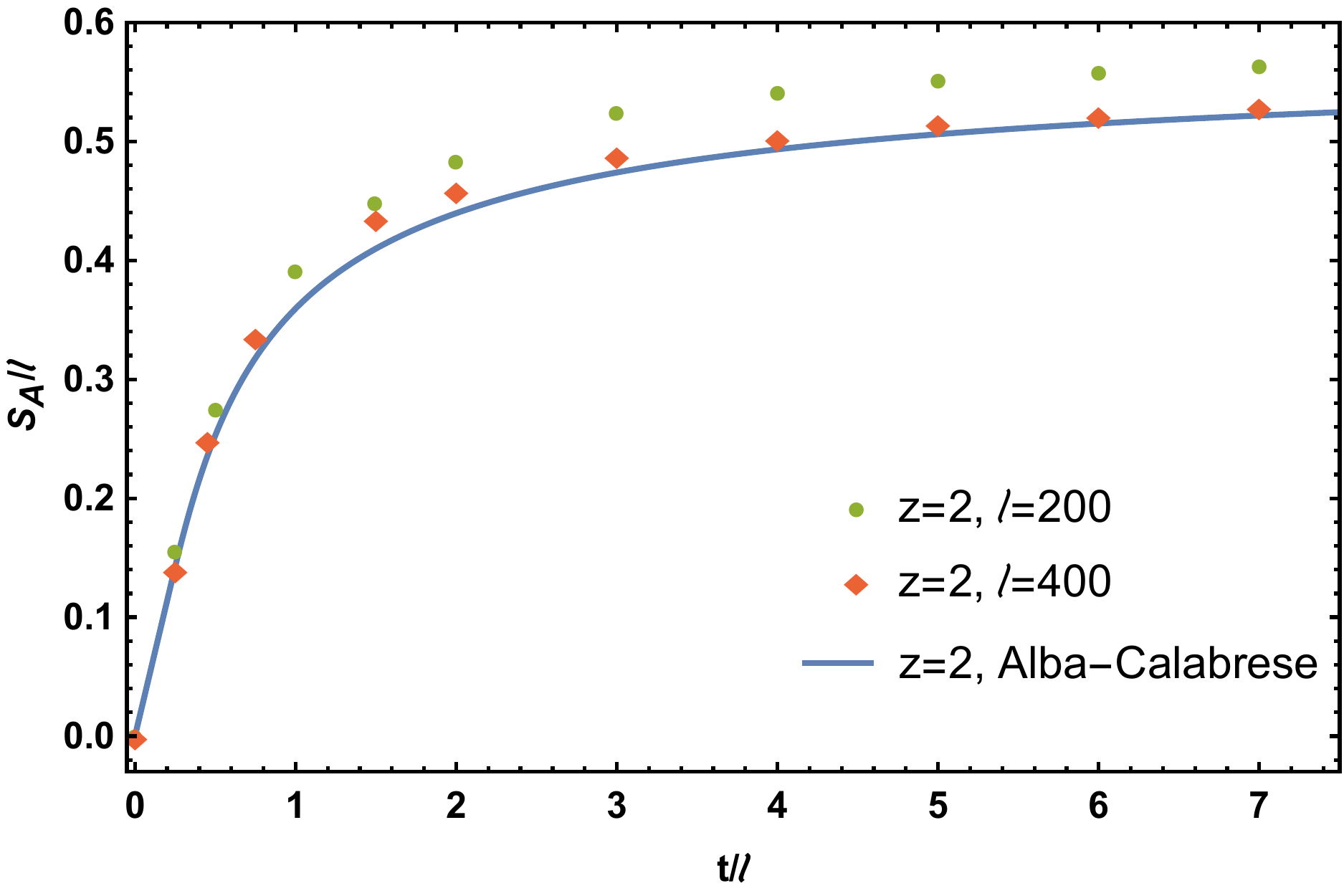}
\hspace{1mm}
\includegraphics[scale=0.28]{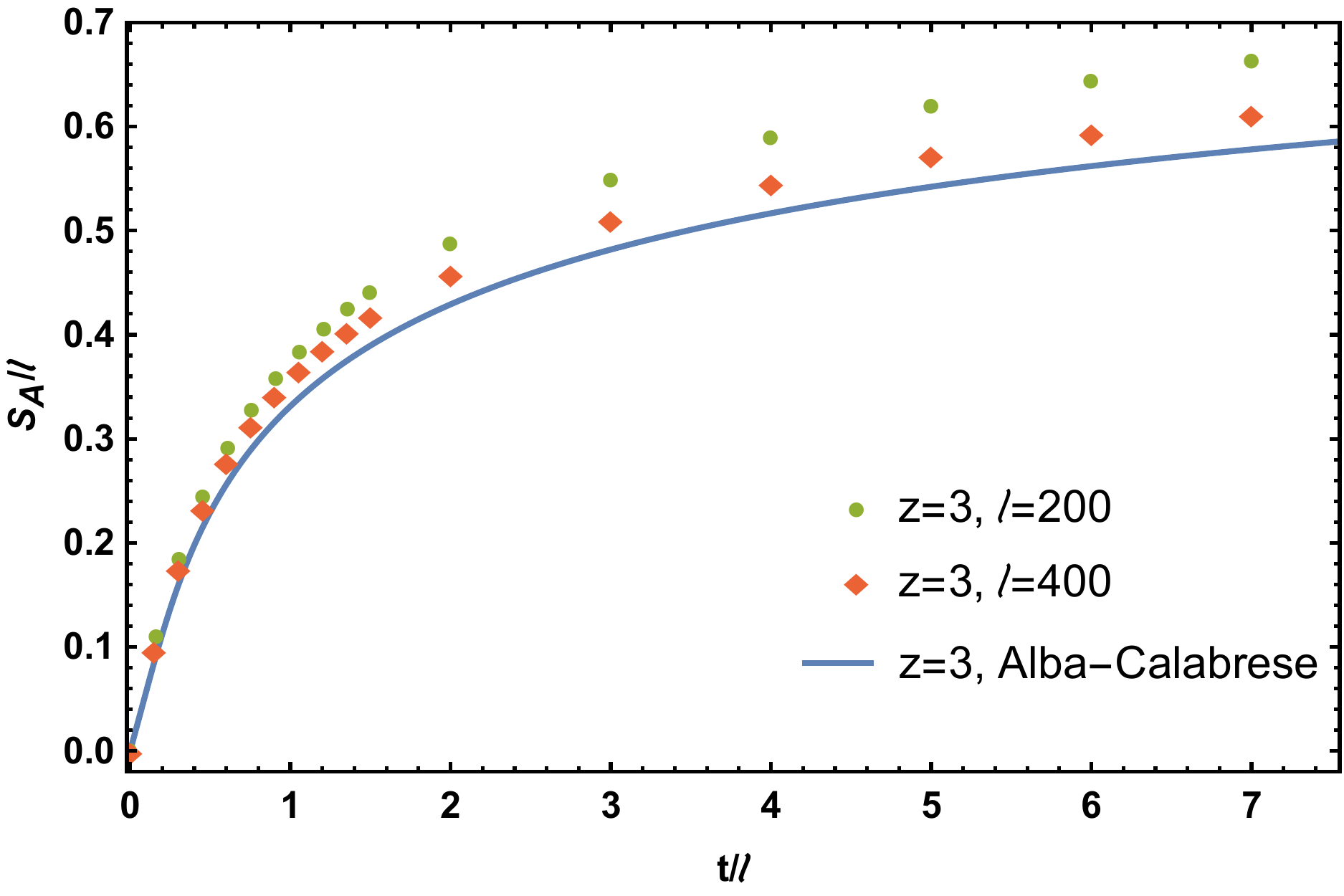}
\end{center}
\caption{Analytic versus numerical results for the evolution of EE in Lifshitz harmonic lattice. In the left panel we have set $m_0=1$ and $m=2$ and in the middle ($z=2$) and right ($z=3$) panels we have set $m_0=1$ and $m=0$. In the left panel there is perfect agreement between Alba-Calabrese quasi-particle picture and numerics. In the middle panel one can see that due to the zero mode effect even for $\ell=400$ numerics deviate from Alba-Calabrese picture around the offset of tortoise saturation regime. The same is correct for higher values of $z$ although one must wait much longer to see this.}
\label{fig:albaz}
\end{figure}

\item For $z<1$, $v_g^{\rm max}$ becomes pure imaginary. Based on this we conclude that there may not be a self-consistent analytical continuation for this model for $z<1$, although our results show that there should be such a continuation for non-integer $z>1$. Based on this we have not considered this range of parameter space of these theories.\footnote{Entanglement dynamics in some long-range models which resembles our mode for $0<z<1$ has been studied in \cite{Nezhadhaghighi:2014pwa, Rajabpour:2014osa}.} It is also worth to mention that due to null energy condition, the same constraint on the dynamical exponent arises in the holographic theories with Lifshitz scaling symmetry \cite{Kachru:2008yh}.

\item In figure \ref{fig:vmaxz} we have plotted $v_g^{\rm max}$ both as a function of the dynamical exponent ($z>1$) and mass parameter. Obviously there is no divergence in the $m\to0$ limit but as expected it diverges as $z\to\infty$.  
\end{itemize}

\subsection{Numerical Results: A Zero Mode Analysis}
In this section we present numerical results for propagation of entanglement after a mass quench in Lifshitz theories. We consider a mass quench while the dynamical exponent as another parameter in the dispersion relation is held fix. We consider two family of quenches which basically differ in the post-quench state. In both families the state prior to the quench is the vacuum state of an infinite Lifshitz harmonic lattice with parameters $(m_0,z)$ where $m_0\neq 0$. The post-quench state is again the vacuum state of an infinite Lifshitz harmonic lattice with parameters changed to $(m,z)$. By two families we mean the post-quench state is either chosen to be massless which the system has Lifshitz scaling symmetry or massive which does not have such a symmetry.

In figure \ref{fig:albaz} we present numerical results regarding to both families and compare them with the quasi-particle picture. In the left panel which the post quench mass is finite one can see perfect agreement between the quasi-particle picture and our numerical results. In the middle and right panel we have shown results for the case where the post-quench mass vanishes. On can see that in these cases since the zero mode effect become more important specifically after the short linear growth regime is finished, there is a deviation from the quasi-particle picture although the deviation is suppressed as one gets closer to the thermodynamic limit. The latter gets much harder to reach as the dynamical exponent is increased because of what we have explained previously about the occupation number of tortoise modes (see figure \ref{fig:nksk}). 

\begin{figure}
\begin{center}
\includegraphics[scale=0.30]{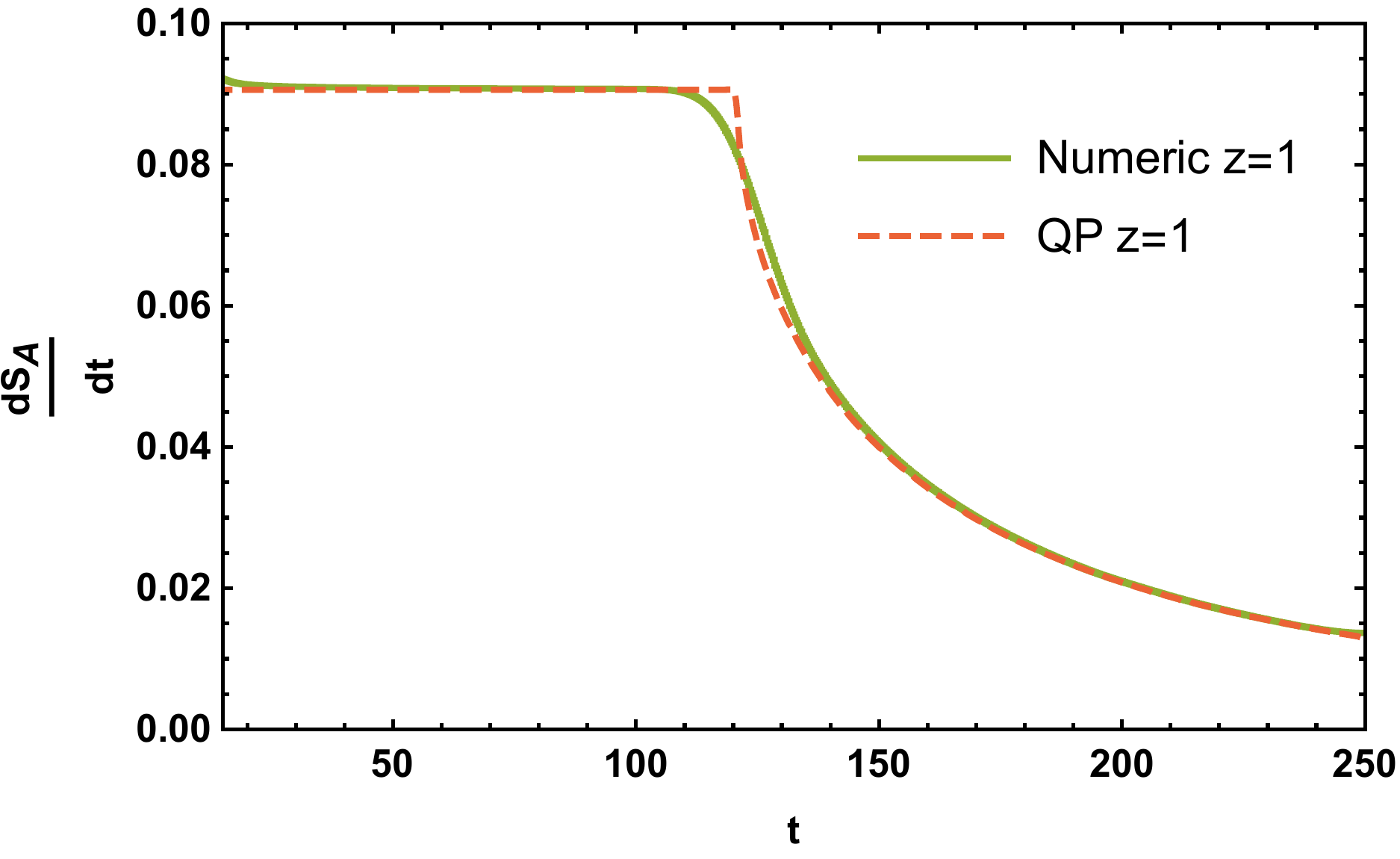}
\includegraphics[scale=0.30]{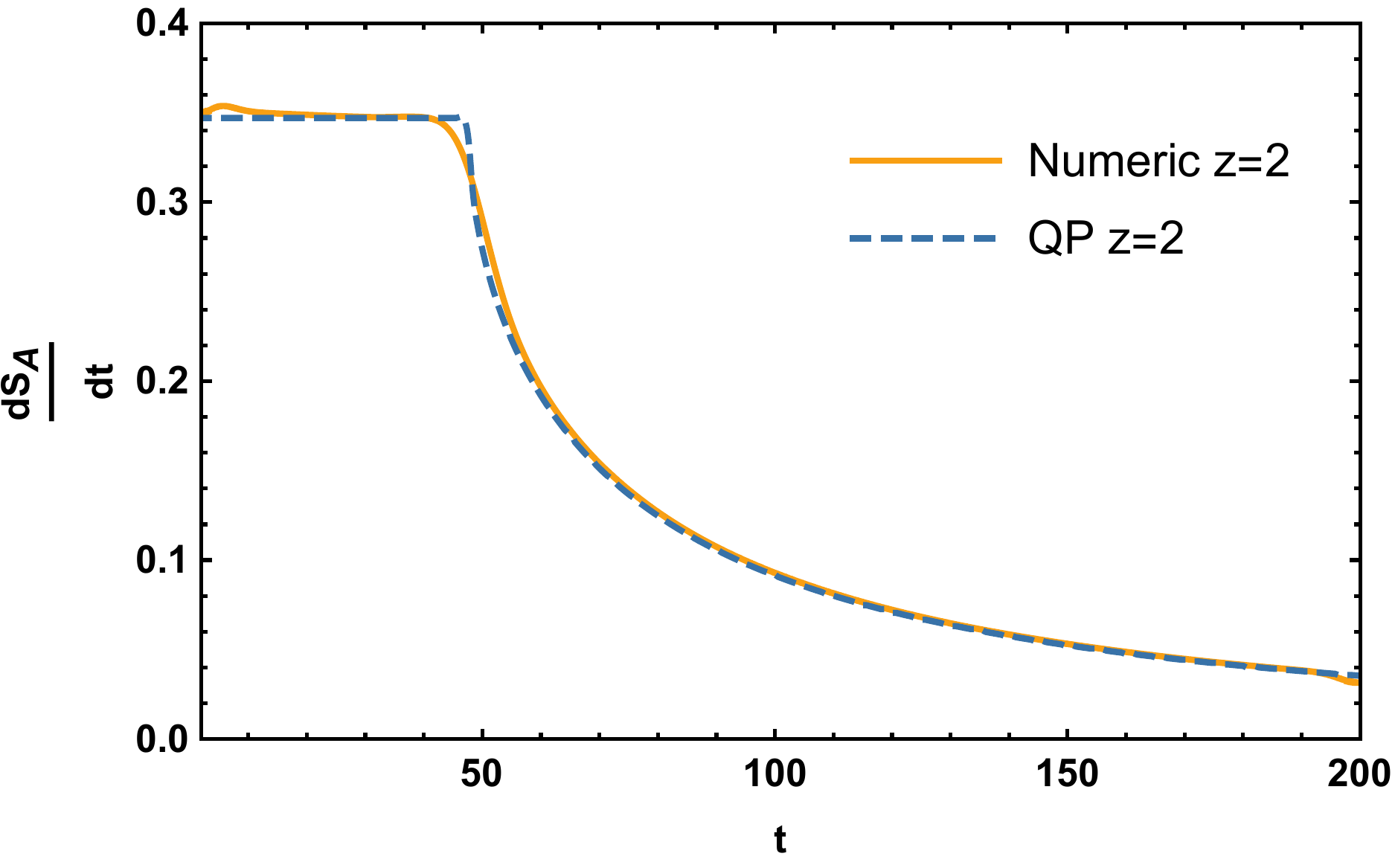}
\includegraphics[scale=0.30]{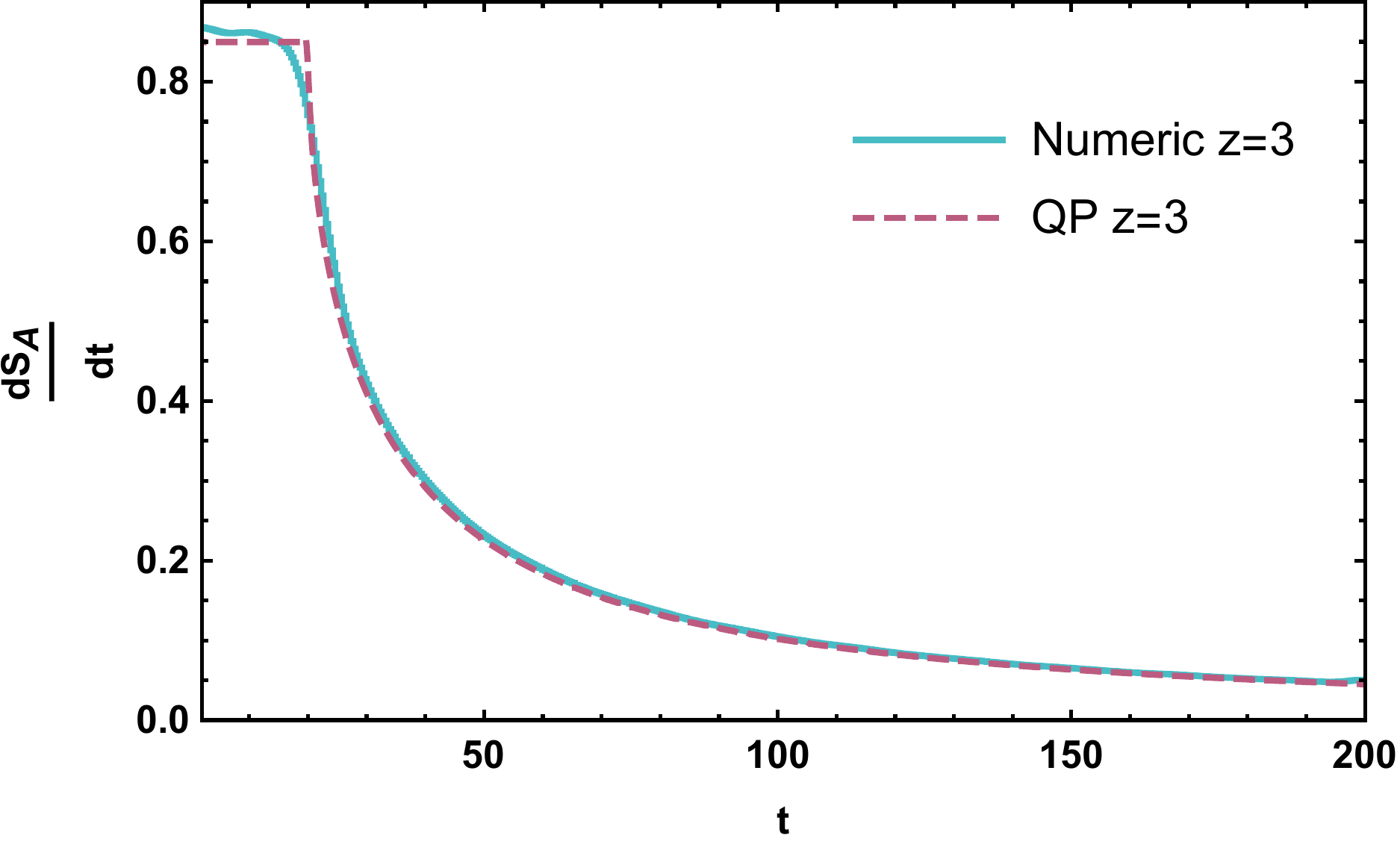}
\end{center}
\caption{Numerical versus analytic results for time evolution of EE for $z=1,2,3$. Here we set $m_0=1$ and $m=2$. There is a very good matching between quasi-particle analytic results and numerics. The small difference at the early times is due to numerical instability and the mild transition in numerical results is due to the zero mode which is not captured by the quasi-particle picture. In these plots we have set the subregion size to be 100 in units of lattice spacing.}
\label{fig:m12}
\end{figure}

In these curves one can see that the larger the value of the dynamical exponent is, the faster EE grows in time. Thus the saturation value of EE is an increasing function of $z$ as expected. This behaviour is expected due to the enhancement of spatial correlations for larger values of the dynamical exponent (for a detailed discussion on the relation between the strength of spatial correlation and $z$ see \cite{MohammadiMozaffar:2017nri}). We have checked this for higher values of $z$ up to $z=5$ but we do not present the results here since the curves have the same qualitative behaviour in different scales. We believe that the physics is perfectly captured by these values of $z$. Also it is notable that we have not studied $z>5$, since in order to interpret the results in the thermodynamic limit and avoid lattice effects for these higher values of $z$, one needs to consider much larger subregions which requires a much higher numerical cost.

\begin{figure}
\begin{center}
\includegraphics[scale=0.3]{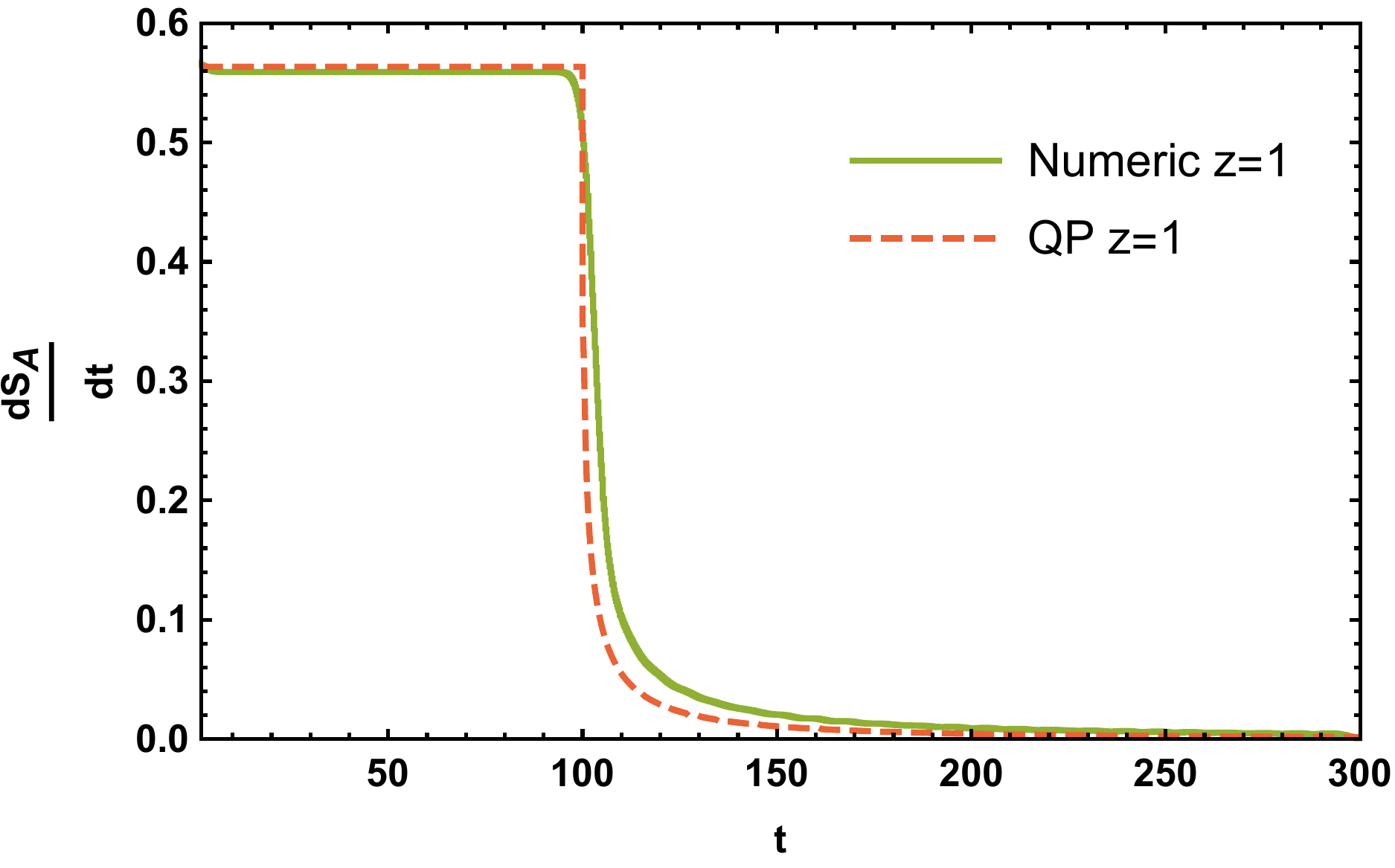}
\includegraphics[scale=0.3]{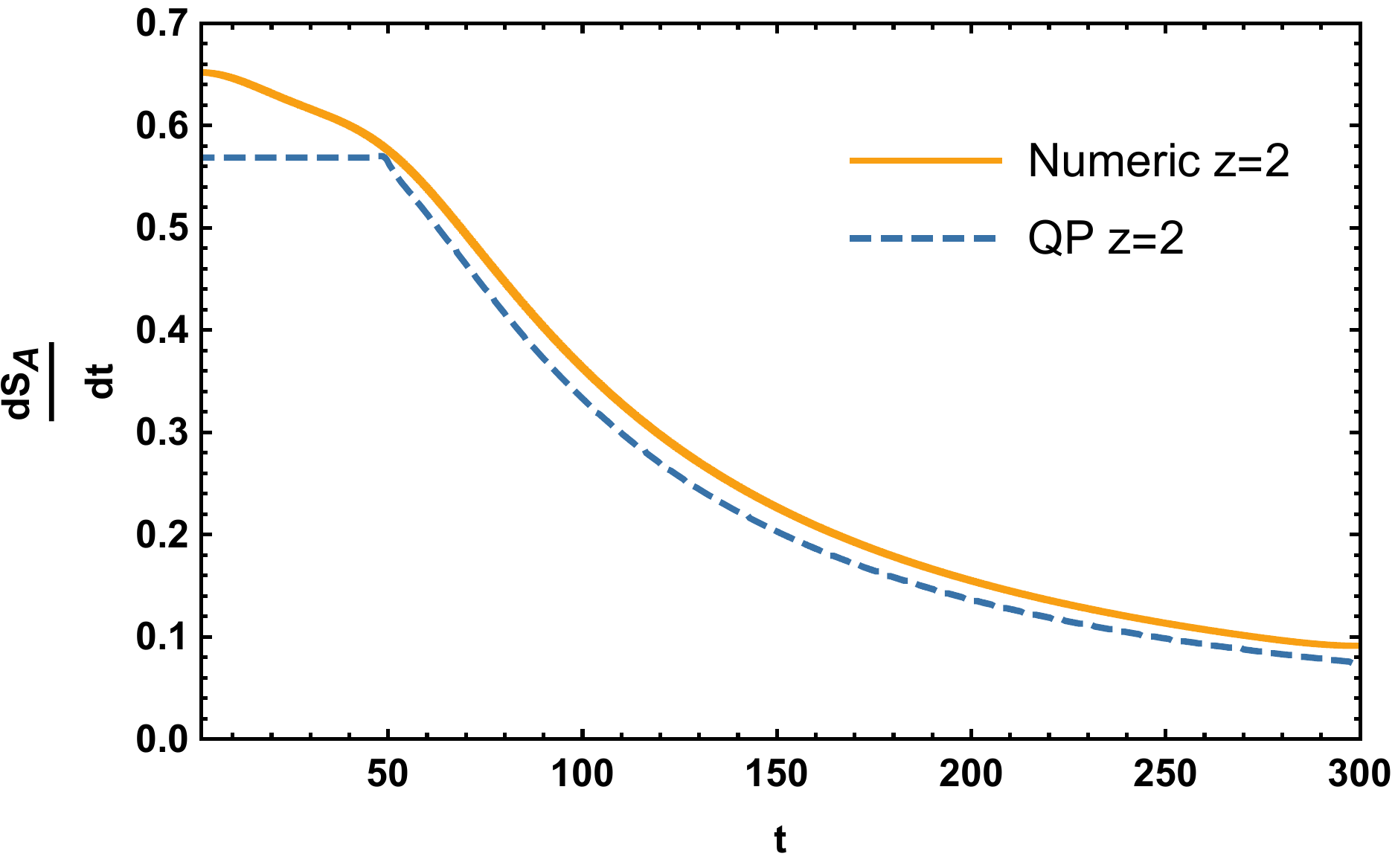}
\includegraphics[scale=0.3]{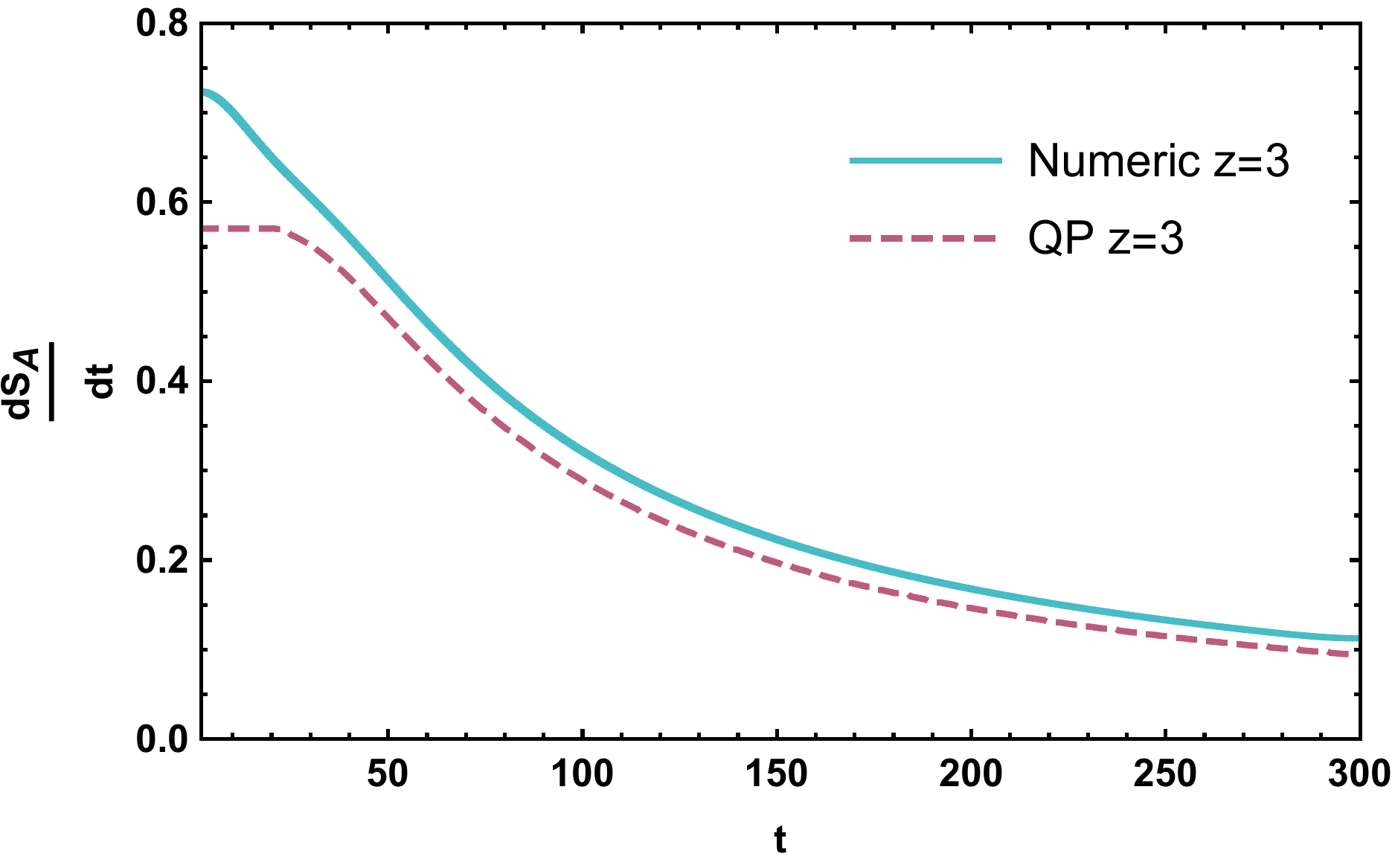}
\end{center}
\caption{Numerical versus analytic results for time evolution of EE for $z=1,2,3$. Here we set $m_0=1$ and $m=10^{-6}$. The matching between quasi-particle analytic results and numerics is not as good as the $m_0=1$ and $m=2$ case. In this case which tortoise modes are highly occupied, the thermodynamic it is much harder to numerically reach the thermodynamic limit. In these plots we have set the subregion size to be 200 in units of lattice spacing.
In this case the difference between numerics and analytical results at the early times is both due to highly occupation of tortoise modes and also numerical instability.}
\label{fig:m10}
\end{figure}

In figure \ref{fig:m12} we focus on numerical results regarding to the first family which is a mass quench from $(m_0=1,z)$ to $(m=2,z)$ for $z=1,2,3$. In figure \ref{fig:m10} we do the same analysis for the second family that is a mass quench from $(m_0=1,z)$ to $(m=0,z)$ for the same values of dynamical exponent. In both figures \ref{fig:m12} and \ref{fig:m10} we have presented the time derivative of EE. In figure \ref{fig:albaz} it is not easy to distinguish between different growth regimes specifically in the second family while this can be clearly seen in  figure \ref{fig:m12}. We would say that figure \ref{fig:m12}, \ref{fig:m10} carry the most important results of this paper since all essential features of propagation of entanglement in these theories can be extracted from them.

There is an initial rapid growth regime which rapidly disappears and is very hard to follow from these cures. During the rapid growth regime the system has not even reached a local-equilibrium and the scaling of entropy is expected to be understood from the only well-defined physical quantity during this regime which is the energy density of the system. We will not present a careful study of this regime here but will make some comments on the $z$-dependent scaling of EE in this regime in the discussion section.

The system rapidly enters its main growth regime which in general longs up to a certain point (see figure \ref{fig:m12} and figure \ref{fig:m10}). After this point a tortoise saturation regime starts which in principle carries on up to infinite time. Both families share this property.  

In general one can obviously mention that in our first family which is the generic case of these theories, the prediction of quasi-particle picture applies here perfectly (figure \ref{fig:m12}) while in the second family which the contribution of the tortoise modes is increased due to the increase in their occupation number there is a serious non-concurrence with the quasi-particle picture (figure \ref{fig:m10}). In the following subsections we will interpret these results and argue that these results can be understood if there is an effective notion of causality in these theories while this is totally non-trivial specially from the field theoretic point of view.

\subsection{Physical Interpretation}\label{sec:physicalInter}

The quasi-particle picture which works perfectly in many cases including 2$d$ CFT and other massive 2$d$ theories also offers a nice understanding of entanglement propagation in Lifshitz theories. Due to our analysis in section \ref{sec:EEz} there is a large spectrum of quasi-particles with different group velocities which are responsible for propagation of entanglement.

The simplest case understood by this picture was the case of 2$d$ CFT which there was a monochromatic quasi-particle with $v_g=1$. In this case the sharp transition from linear growth to saturation regime is totally clear. At the instance that the quench happens, a pair of monochromatic quasi-particles start to propagate back-to-back to each other at all spatial points and they can pass through the entangling region. At early times those pairs initiated from the spatial points near the boundary of the entangling region have one mate inside the region and the other outside. These pairs are responsible for the linear growth. After a while proportional to the length of the region, the phenomena of ingoing and outgoing of quasi-     particles through the entangling region equilibrates and thus EE saturates.     

In the case of massive theories where there exists a spectrum of quasi-particles with different group velocities, the situation is a bit more complex. One should consider the behaviour of different types of quasi-particles to understand the propagation of entanglement. The group velocity in this case is constrained to $0\le v_g \le 1$. All quasi-particles propagate inside the light-cone. Those with maximum group velocity propagate on the null directions and the zero modes depending on the value of the mass propagate extremely slow. In presence of these extremely slow modes, the above scenario still works for all types of quasi-particles. After the effect of the fast quasi-particles, specifically the one with maximum group velocity is equilibrated, the slow modes take over the role. Due to existence of these modes there always exists pairs which one mate is inside the entangling region and the other mate is outside. Although the number of these pairs is decreasing in time, but the decrease rate is infinitely slow thus the tortoise saturation regime extend over infinite time.

\begin{figure}
\begin{center}
\includegraphics[scale=0.35]{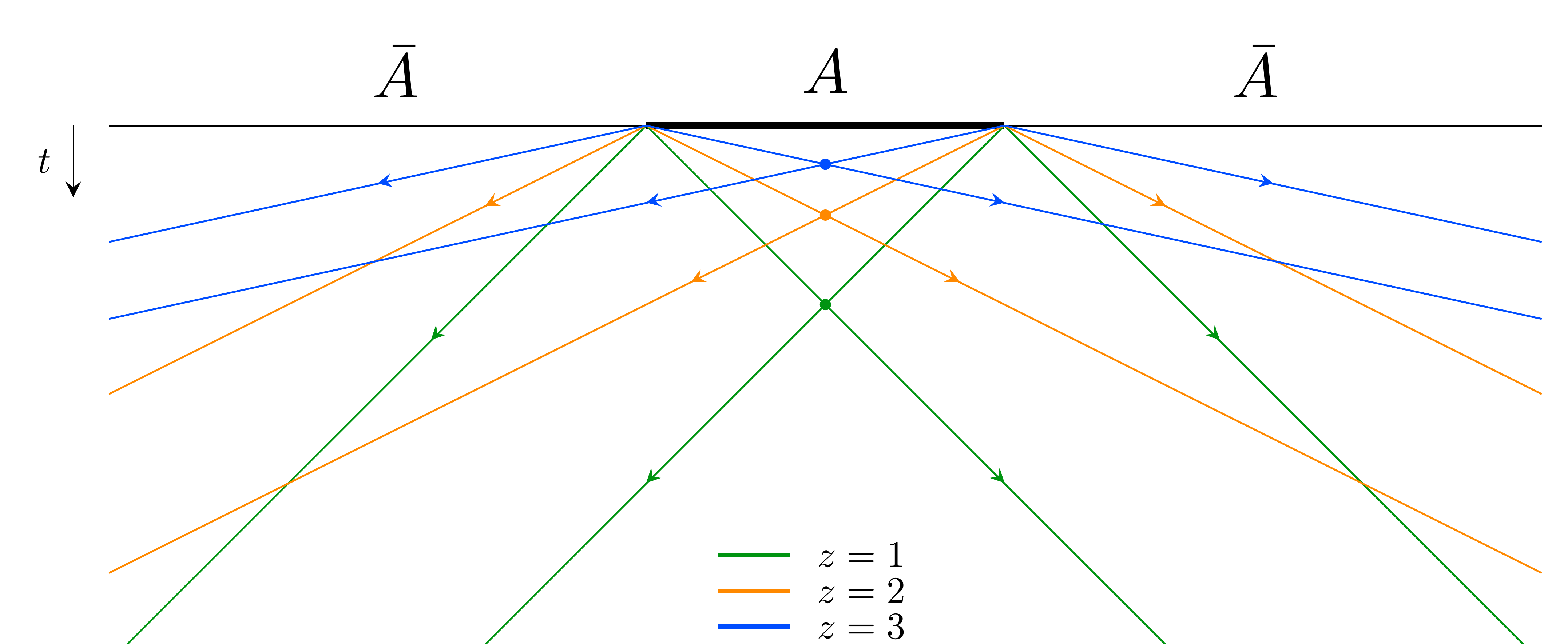}
\end{center}
\caption{Here $A$ is the entangling region and $\bar{A}$ is the complement. The widen light-cones for different values of dynamical exponent are shown. The green lines are the standard (Lorentzian) light rays with $v_g^{\text{max}}(1)=1$ and the orange and blue ones correspond to higher dynamical exponents which the cone gets wider. The specified points in the middle of the region along the time direction are $\{t^{\text{max}}_s(1),t^{\text{max}}_s(2),t^{\text{max}}_s(3)\}$. As $z$ gets larger, $t^{\text{max}}_s(z)\to 0$ and the boundaries of the widen light-cone tend to lie on the $t=0$ axis.}
\label{fig:qpz}
\end{figure}

The very interesting thing about Lifshitz theories is that even in the massless case, there exists a spectrum of quasi-particles with $0\le v_g(z) \le v_g^{\text{max}}(z)$, where $v_g(z)$ is given in Eq.\eqref{vgz} and $v_g^{\text{max}}(z)$ is given in Eq.\eqref{vgm0max} and $v_g^{\text{max}}(1)=1$. The quasi-particles propagate inside a \textit{widen light-cone} which its maximum velocity rays are given by $v_g^{\text{max}}$. In figure \ref{fig:qpz} we have shown this behaviour for an entangling region for values of $z=\{1,2,3\}$. The slope of the blue, orange and green rays are respectively given by $\{\pm v_g^{\text{max}}(1),\pm v_g^{\text{max}}(2),\pm v_g^{\text{max}}(3)\}$. The three coloured points are the saturation times found from the fastest monochromatic quasi-particle. For a single interval entangling region, the saturation time for a monochromatic quasi-particle is found to be defined as the intersection of the world-line of two of these quasi-particles starting from the end points of the region. This is simply found to be $t^{\text{max}}_s(z)=\ell/(2v_g^{\text{max}}(z))$. 

The interesting point here is that in Lifshitz theories there still exists an effective notion of \textit{saturation time} which we denote by $t^*_s(z)$. The saturation time is defined to be the instance that evolution of EE experiences a sudden transition from the main growth regime to the tortoise saturation regime. This notion is also well-defined for $z=1$ case, which is specifically important when the system is massive and a spectrum of quasi-particles are propagating around. 

Our numerical data hints to interesting physics in Lifshitz theories. The picture is that before EE enters the tortoise saturation regime, the monochromatic quasi-particle with $v_g^{\text{max}}(z)$ dominates the entanglement propagation. This the physical reason why figure \ref{fig:qpz} is a very good approximate for the physics of entanglement propagation in the main growth regime. In other words it is clear that in principle $t^{\text{max}}_s(z)$ is not the same as the effective saturation time we have defined above by $t^*_s(z)$. But here what we find is that these two are actually the same.   

This picture clearly shows that in presence of tortoise modes the precise saturation time goes to infinity, meanwhile the effective saturation time $t^*_s(z)$ goes to zero (following the green, orange and blue points in figure \ref{fig:qpz} toward the $t=0$ axis). In other words as the dynamical exponent increases, the dominant regime in time is the tortoise saturation regime.      

The other interesting behaviour which is also understandable with the quasi-particle picture is comparing the behaviour of EE for different values of $z$ in the tortoise saturation. As we have argued the tortoise modes are mainly responsible for this behaviour. We have shown in \ref{subsec:analytic} that the occupation number increases while $z$ is increased. This is the reason why the tortoise saturation regime becomes slower for higher values of $z$. The EE saturates at infinite time for $z>1$, and these infinite times comparing to each other is larger for higher values of $z$.  

\subsection{Lattice versus Continuum}\label{sec:LatvCon}
An interesting question which was one of the main motivations for this study is how propagation of entanglement is related to the causality structure of the theory. It is well-known that propagation of entanglement is governed by the dispersion relation (spectrum of the quasi-particles) and the causality structure of the theory of interest. This was clearly understood at least in the case of 2$d$ CFTs and massive scalar theory. On the other hand the picture is also consistent with the lattice version of massive scalar theory, i.e. the harmonic lattice.

\begin{figure}
\begin{center}
\includegraphics[scale=0.30]{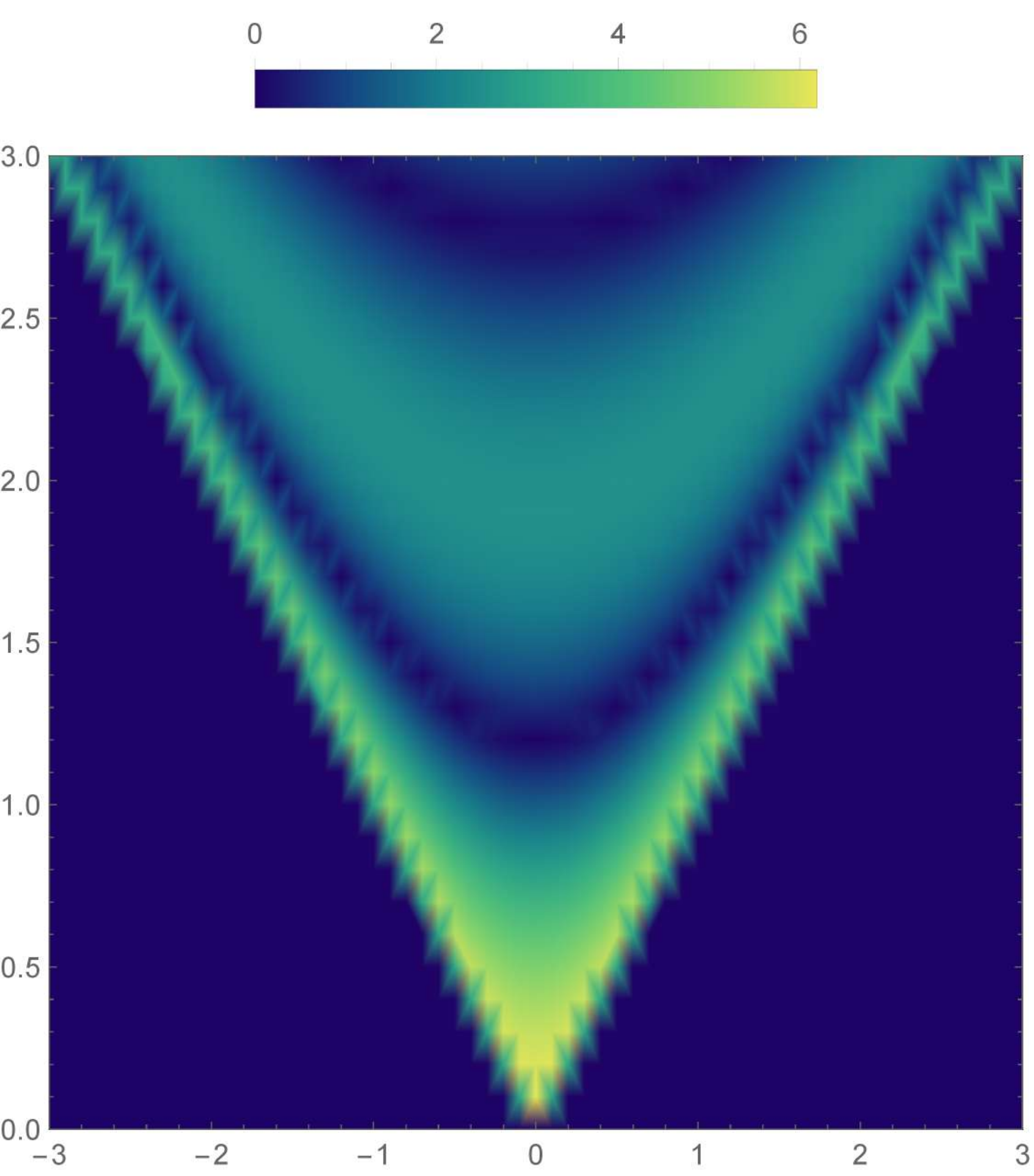}
\includegraphics[scale=0.30]{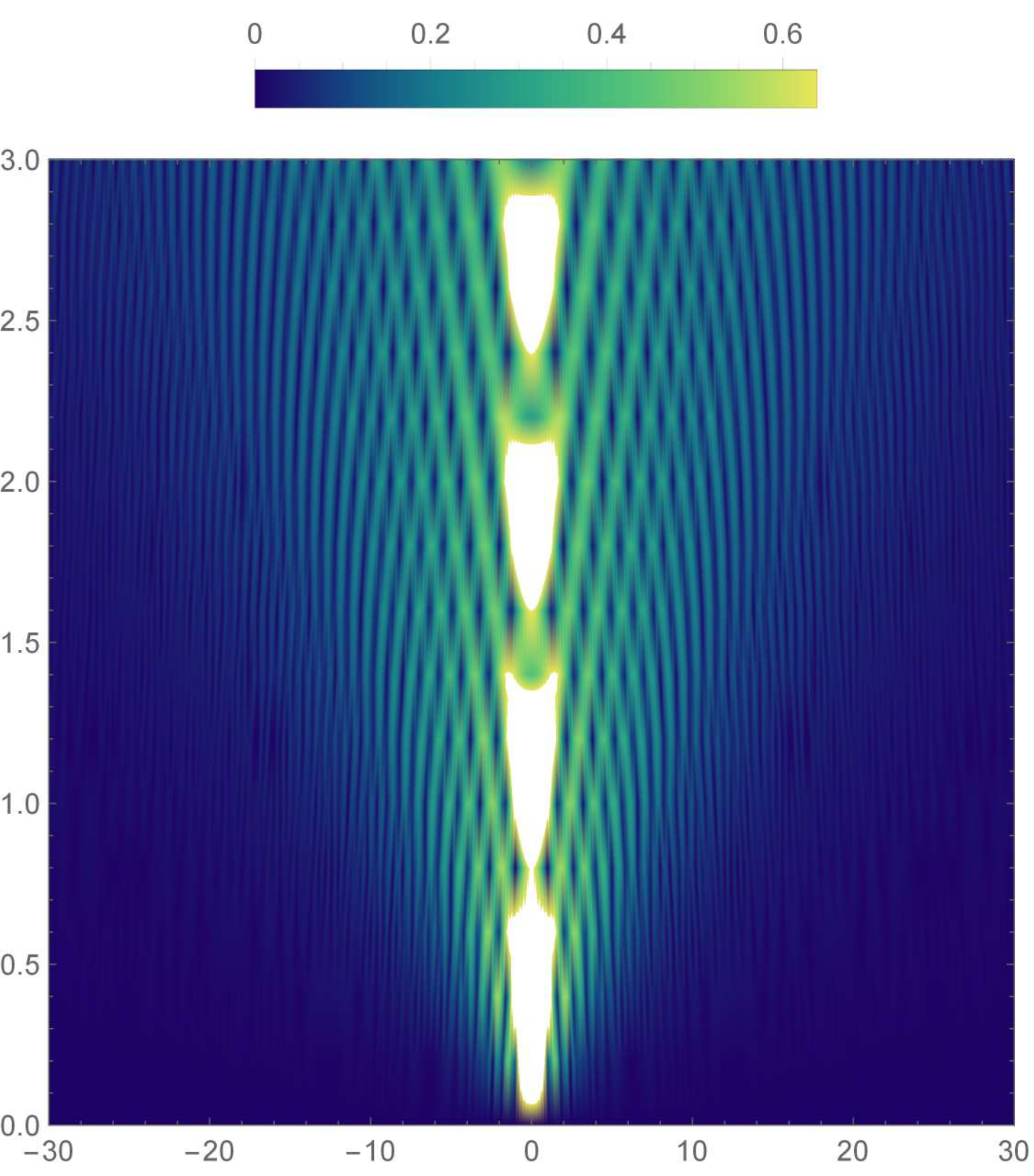}
\includegraphics[scale=0.30]{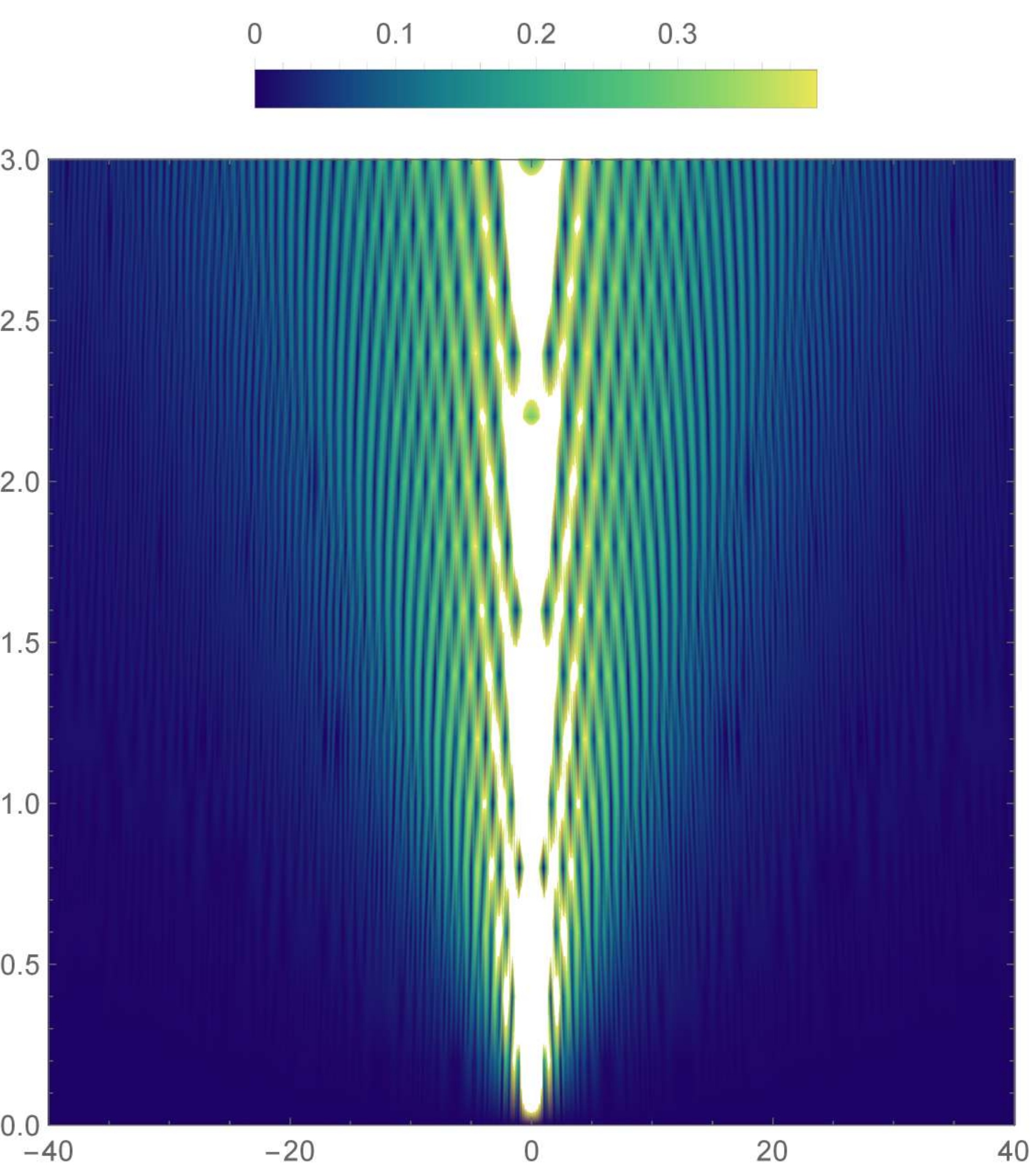}
\end{center}
\caption{Density plot for the vacuum expectation value of $[\phi(0,0),\phi(x,t)]$ in 2$d$ continuous theory. The plots refer to $z=1,2,3$, from left to right respectively. The horizontal axes is $x$ and the vertical axes is $t$. In all plots we have set $m=2$ and the UV cut-off to be $\Lambda=10^3$.}
\label{fig:density}
\end{figure}

Lets more precisely compare what is going on in our lattice model and how these results are supported by a simple analysis in the continuum limit. The dispersion relation and group velocity of propagating modes in this theory is given by
\bea
\omega(k)=\sqrt{m^{2z}+k^{2z}}\;\;\;\;,\;\;\;\;v_g(k)=\frac{z k^{2z-1}}{\sqrt{m^{2z}+k^{2z}}}.
\eea
Clearly this group velocity does not have an extremum at least for $z>1$. This makes it highly counter-intuitive to have a causal structure in these theories. On the other hand our results which are strongly supported by the quasi-particle picture (modulo the states which the zero modes are highly occupied) show that there is at least an effective notion of causal structure due to propagation of entanglement in these theories. It is worth to note that we are comparing the discrete model in  the thermodynamic limit with the continuum theory.

This is actually what is well-known in the context of many-body systems by the Lieb-Robinson bound \cite{LR}. There are several well known models which in the continuum correspond to local theories which do not have a Lorentzian causal structure but due to locality the correlations between points decay exponentially with their distance and there is a Lieb-Robinson velocity defined by the bound which information cannot spread faster it. 

To be more concrete in the field theory language, there should be a light-cone like structure in the theory which measurements inside the cone should not be affected from outside the cone and vice versa. In the language of our theory in the simple case of $d=1$, this means that the simplest thing is to look at the commutator of fields
\begin{align}
\begin{split}
\langle[\phi(0), \phi(y)]\rangle
&=\int \frac{dk}{2\pi}\frac{1}{2\omega(k)}\left(e^{i\omega(k)t-iky}-e^{-i\omega(k)t+iky}\right),
\end{split}
\end{align}
at different space-time points. In the Lorentzian case this is going to vanish for any spacelike separated points. In the case of Lifshitz scalar theory we present a numerical study of this quantity for different values of dynamical exponent and mass parameter in figure \ref{fig:density}. It can be shown from the plots that there exists an effective widen cone that the commutator vanishes outside it. This is in agreement from what we expect from Lieb-Robinson bound in the lattice version and also with what we have found from the behaviour of entanglement entropy. We postpone a more concrete study of causality in these theories to \cite{toAppear1}.\footnote{For a related study of correlation functions in Lifshitz theories with $d=z$ see \cite{Keranen:2016ija}.}

\section{Conclusions and Discussions}
We have studied the relaxation process of quenched states in free Lifshitz scalar theories to generalized Gibbs ensemble. Our study was mainly in 2$d$, which is the simplest case to utilize the quasi-particle picture to understand the process physically. A wide range of dynamical exponents is studied although we have presented results for a few number of them which was enough to focus on the physical picture. An important feature of these theories is a momentum-dependent group velocity of its propagating modes.

We have utilized a discrete version of these theories, i.e., Lifshitz harmonic lattice models introduced in \cite{MohammadiMozaffar:2017nri}. Different regimes of growth of entanglement entropy in these theories after a sudden quench is studied. Our results are mainly captured by an improved version of the quasi-particle picture introduced in \cite{Alba:2017lpnas}. In the following we would like to summarize our main results:

\begin{itemize}
\item
Comparing two specific values of dynamical exponent, say $z_1$ and $z_2$ which $z_2>z_1$, as expected from the enhancement of equal-time correlation functions we have shown that the value of EE is larger for $z_2$. The growth rate of EE is also grater for the $z_2$ case, except in a short period of time after $z_2$ has entered the tortoise saturation regime while $z_1$ is still linearly increasing. This short period is expected due to two different phenomena. On one hand as the dynamical exponent increases the linear regime is shortened, and on the other hand the occupation number of slow modes is increased. Thus one would expect that the growth rate of $z_1$ is larger than $z_2$ during the end of the late linear regime of $z_1$.  

\item
We have shown that the larger the dynamical exponent is, the slower EE saturates. Actually exact saturation for these theories (and for any theory admitting a pile of slow modes) is postponed to infinity. For larger values of dynamical exponent, this process becomes slower and slower. This effect is due to the increase of occupation number of slow modes with the dynamical exponent. Based on this, even in the scale invariant case, i.e., the mass parameter of the post-quench Hamiltonian vanishes, no sudden saturation happens in these theories in contrast with scale invariant 2$d$ relativistic theories.   

\item
We have shown that except the cases which the contribution of tortoise modes becomes dominant (in other words the occupation number of these modes is much larger than that for fast modes), the Alba-Calabrese quasi-particle picture works perfectly in these theories. We have shown this for EE and its time derivative for mass quenches. Thanks to Alba-Calabrese, we have an analytic expression for propagation of entanglement in these theories. As a byproduct we have worked out the analytical time dependence of EE during the tortoise saturation regime for $z=1$ and $z=2$. 

\item
Our result which was explained in sections \ref{sec:physicalInter} and \ref{sec:LatvCon} are showing that the propagation of entanglement is well understood with the existence of an effective notion of widen light-cone in these theories. The larger the dynamical exponent, the wider the light-cone is. Our study is a strong, although still indirect, representative of existence of at least an effective causal structure in Lifshitz theories. Of course from the dispersion relation of these theories this is not obvious at all. We postpone a rigorous study of causal structure in these theories to a later work \cite{toAppear1}. 

\item We would like to compare our results with holographic studies of evolution of EE in Lifshitz space times \cite{Alishahiha:2014cwa, Fonda:2014ula}. This is mainly studied by considering a Vaidya geometry with an asymptotically Lifshitz spacetime leading to a finite saturation time. But we have shown that based on the existence of tortoise modes the tortoise saturation regime is prolonged and due to that the saturation time goes to infinity. This is another sign for non-robustness of considering asymptotically Lifshitz geometries in a relativistic theory as a dual to a state in a Lifshitz-type theory.\footnote{For a similar analogy see \cite{Cheyne:2017bis} where the authors have argued that for a correct entanglement wedge reconstruction of Lifshitz spacetime, one needs to go beyond relativistic gravity theories, for instance Horava-Lifshit gravity.}

\end{itemize}

There are some other results/comments regarding to our study which we would like to mention in the following:

We think that a interesting question which becomes more significant after this work is how is it possible to generalize the quasi-particle picture to be able to capture the strange effects of the tortoise modes in the dynamics?

In this work we have considered a sudden quench since it is the simplest framework to study propagation of entanglement turning off other interesting features of the theory and only focusing on propagation of entanglement. The more general question is what are the critical exponents of these theories under a Kibble-Zurek phase transition which our study is a fast quench limit of that more general setup (for similar studies in Lorentzian theories see \cite{Buchel:2013gba, Das:2016lla, Caputa:2017ixa, Das:2017sgp}). We postpone reporting results regarding to this question to a feature work \cite{toAppear2}.   

The numerical analysis of the early time rapid growth in these theories show that the growth of EE in this regime seems to be have a very good fit with $S\sim t^{1+\frac{1}{z}}$. This behaviour was previously found in the holographic context \cite{Alishahiha:2014cwa, Fonda:2014ula}. This kind of scaling with $t$ seems a bit confusing. The reason is that one would expect EE in the very early region after a sudden quench, which even no local equilibrium state is reached, to scale with $\mathcal{E}\cdot \mathcal{A}\cdot t^\alpha$, where $\mathcal{E}$ the energy density of the system, $\mathcal{A}$ is the area of the entangling region and $\alpha$ is fixed by dimensional analysis which in a Lifshitz theory terns out to be $1+z$. This behaviour is clearly expected to be the case for any number of spatial dimensions. It would be interesting to figure out what is the correct scaling of entanglement in this regime.

Another interesting future direction of this work would be considering a new family of quenches regarding to the other parameter in the Hamiltonian. The dynamical exponent is a parameter in the dispersion relation very similar to the mass parameter. It would be interesting to study the pattern of EE after a dynamical exponent quench specifically from a scale invariant theory ($m_0=0$ and $z=1$) to another scale invariant theory  ($m_0=0$ and $z>1$).

\subsection*{Acknowledgements}
We would like to thank Alex Belin, Diptarka Das, Michal Heller, Christoph Herzog for fruitful discussions. We would like to also thank Bruno Nachtergaele for correspondence. We are grateful to Pasquale Calabrese and Tadashi Takayanagi for their comments on a draft of this manuscript.  
AM thanks the organizers of ``Workshop on AQFT, Modular Techniques, and Renyi Entropy" held at AEI in Potsdam where these results where first presented.
AM is supported by Alexander-von-Humboldt Foundation through a postdoctoral fellowship.

\appendix
\section{Time Dependent Correlator Method}
We are now interested in a case that the frequency of the Hamiltonian is suddenly changed from its initial value $\omega_{0,\mathbf{k}}$ to $\omega_{\mathbf{k}}$. After this sudden change the vacuum state of the initial Hamiltonian evolves unitarily with the new Hamiltonian. If we denote the vacuum state of the initial state as $|0\rangle$ we need to compute the following correlators in order to study the time evolution of entanglement and Renyi entropies.

\begin{align}
\begin{split}
X_{\mathbf{i}\mathbf{j}}(t)&=\langle0|\phi_{\mathbf{i}}(t)\phi_{\mathbf{j}}(t)|0\rangle,\\
P_{\mathbf{i}\mathbf{j}}(t)&=\langle0|\pi_{\mathbf{i}}(t)\pi_{\mathbf{j}}(t)|0\rangle,\\
R_{\mathbf{i}\mathbf{j}}(t)&=\langle0|\phi_{\mathbf{i}}(t)\pi_{\mathbf{j}}(t)|0\rangle,
\end{split}
\end{align}
where
$$\phi_{\mathbf{i}}(t)=e^{i H(\omega_{\mathbf{k}})t}\phi_{\mathbf{i}}(0)	e^{-i H(\omega_{\mathbf{k}})t}
\;\;\;\;\;,\;\;\;\;\;
\pi_{\mathbf{i}}(t)=e^{i H(\omega_{\mathbf{k}})t}\pi_{\mathbf{i}}(0)	e^{-i H(\omega_{\mathbf{k}})t}.
$$
The explicit form of these correlators is given by
\begin{align}\label{eq:correlators}
\begin{split}
X_{\mathbf{i}\mathbf{j}}(t)&=
\frac{1}{2}\prod_{r=1}^d \frac{1}{N_{x_r}}\sum_{k_r=0}^{N_{x_r}}\mathbb{X}_{\mathbf{k}}
\cos\left(\frac{2\pi(i_r-j_r)k_r}{N_{x_r}}\right),\\
P_{\mathbf{i}\mathbf{j}}(t)&=
\frac{1}{2}\prod_{r=1}^d \frac{1}{N_{x_r}}\sum_{k_r=0}^{N_{x_r}}\mathbb{P}_{\mathbf{k}}
\cos\left(\frac{2\pi(i_r-j_r)k_r}{N_{x_r}}\right),\\
R_{\mathbf{i}\mathbf{j}}(t)&=
\frac{1}{2}\delta_{\mathbf{i},\mathbf{j}}-\frac{1}{2}\prod_{r=1}^d \frac{1}{N_{x_r}}\sum_{k_r=0}^{N_{x_r}}\mathbb{R}_{\mathbf{k}}
\cos\left(\frac{2\pi(i_r-j_r)k_r}{N_{x_r}}\right),
\end{split}
\end{align}
where
\begin{align}
\begin{split}
\mathbb{X}_{\mathbf{k}}&=
\frac{1}{\omega_{\mathbf{k}}}\left(\frac{\omega_{\mathbf{k}}}{\omega_{0,\mathbf{k}}}\cos^2\omega_{\mathbf{k}}t+\frac{\omega_{0,\mathbf{k}}}{\omega_{\mathbf{k}}}\sin^2\omega_{\mathbf{k}}t\right)
\\
\mathbb{P}_{\mathbf{k}}&=
\omega_{\mathbf{k}}\left(\frac{\omega_{\mathbf{k}}}{\omega_{0,\mathbf{k}}}\sin^2\omega_{\mathbf{k}}t+\frac{\omega_{0,\mathbf{k}}}{\omega_{\mathbf{k}}}\cos^2\omega_{\mathbf{k}}t\right)
\\
\mathbb{R}_{\mathbf{k}}&=
\left(\frac{\omega_{\mathbf{k}}}{\omega_{0,\mathbf{k}}}-\frac{\omega_{0,\mathbf{k}}}{\omega_{\mathbf{k}}}\right)\sin\omega_{\mathbf{k}}t\cos\omega_{\mathbf{k}}t
\end{split}
\end{align}
It is worth to note that in the numerical calculations of this paper we have used a continuous version of these correlators, that is we have used the integral version of Eq.\eqref{eq:correlators}, that is $N_{x_1}\to\infty$.
Having these we are almost equipped to compute the entropy via the eigenvalues of $i J\cdot \Gamma$ which we denote by $\{\nu_k(t)\}$ where
\be
\Gamma=\begin{pmatrix}
X&R\\R^T&P
\end{pmatrix}
\;\;\;\;\;,\;\;\;\;\;
J=\begin{pmatrix}
0&\mathbb{1}\\-\mathbb{1}&0
\end{pmatrix}.
\ee
The entropies are given by
\begin{align}
S_A&=\sum_{k=1}^{n_A}\left[\left(\nu_k(t)+\frac{1}{2}\right)\log\left(\nu_k(t)+\frac{1}{2}\right)-\left(\nu_k(t)-\frac{1}{2}\right)\log\left(\nu_k(t)-\frac{1}{2}\right)\right],\label{EE}\\
S^{(n)}_A&=\frac{1}{n-1}\sum_{k=1}^{n_A}\log\left[\left(\nu_k(t)+\frac{1}{2}\right)^n-\left(\nu_k(t)-\frac{1}{2}\right)^n\right],
\end{align}
where $n_A$ is the number sites in region $A$.

\end{document}